\documentclass[submission, Phys]{SciPost}

\usepackage[utf8]{inputenc} 
\usepackage[T1]{fontenc} 	
\usepackage[english]{babel} 

\usepackage[bitstream-charter]{mathdesign}
\usepackage{geometry} 		
\usepackage{amsmath} 		
\usepackage{mathtools} 		
\usepackage{float} 			
\usepackage{graphicx} 		
\usepackage{tabularx} 		
\usepackage{booktabs} 		
\usepackage{color, xcolor} 	
\usepackage{pdfpages} 		
\usepackage{extarrows} 		
\usepackage{multirow} 		
\usepackage{multicol} 		
\usepackage{enumitem} 		
\usepackage{xspace} 		
\usepackage{stackrel} 		
\usepackage{tikz} 			
\usepackage{braket} 		
\usepackage{bm} 			
\usepackage{tensor} 		
\usepackage{slashed} 		
\usepackage{siunitx} 		
\usepackage{lastpage} 		
\usepackage{cite} 			
\usepackage[normalem]{ulem} 
\usepackage{fontawesome} 	
\usepackage{tocloft} 		
\usepackage{titlesec} 		
\usepackage{doi} 			
\usepackage{hyperref} 		
\usepackage[most]{tcolorbox} 					
\usepackage[nameinlink, capitalize]{cleveref} 	
\usepackage[nottoc, notlot, notlof]{tocbibind} 	
\usepackage[ruled, vlined]{algorithm2e} 		
\usepackage{makecell}
\usepackage[makeroom]{cancel}
\usepackage{feynmf}
\usetikzlibrary{arrows, shapes.geometric, arrows.meta, shapes, decorations.pathreplacing, fit}

\makeatletter
\def\BState{\State\hskip-\ALG@thistlm}
\makeatother

\makeatletter
\@ifundefined{pdfoutput}{}{\DeclareGraphicsRule{*}{mps}{*}{}}
\makeatother

\makeatletter
\DeclareRobustCommand*{\bfseries}{%
   \not@math@alphabet\bfseries\mathbf
   \fontseries\bfdefault\selectfont
   \boldmath
}
\makeatother

\hypersetup{
	pdftitle={SPINUP},
	pdfauthor={Heimel et al.},
	colorlinks=true, 			
	linkcolor={red!50!black}, 	
	citecolor={blue!50!black}, 	
	urlcolor={blue!80!black} 	
} 

\DeclareSymbolFont{usualmathcal}{OMS}{cmsy}{m}{n}
\DeclareSymbolFontAlphabet{\mathcal}{usualmathcal}

\SetArgSty{textnormal}
\SetKwComment{Comment}{{\small\#}~}{}
\SetCommentSty{mycommfont}

\setitemize{itemsep=0pt, parsep=0pt} 				
\setenumerate{itemsep=0pt, parsep=0pt} 				
\setitemize{itemsep=2pt,topsep=2pt,parsep=0pt,partopsep=0pt,leftmargin=*}
\setenumerate{itemsep=0pt,topsep=2pt,parsep=0pt,partopsep=0pt,labelindent=3pt,leftmargin=*}
\usepackage{amsmath}
 
\usepackage{amsthm} 		
\theoremstyle{definition}
 
\definecolor{red_cb}{HTML}{e41a1c}
\definecolor{blue_cb}{HTML}{377eb8}
\definecolor{green_cb}{HTML}{4daf4a}
\definecolor{purple_cb}{HTML}{984ea3}
\definecolor{orange_cb}{HTML}{ff7f00}

\definecolor{EmeraldGreen}{HTML}{1ea78d}
\definecolor{EnglishRed}{HTML}{b02427}
\hypersetup{colorlinks=true,urlcolor=EmeraldGreen,citecolor=EmeraldGreen,linkcolor=EnglishRed}

\newcommand{\ie}{\text{i.e.}\;}

\newcommand{\kl}{D_\text{KL}}

\newcommand{\psim}{p_\text{sim}}

\newcommand{\pd}{p_\text{data}}

\newcommand{\punf}{p_\text{unfold}}
\newcommand{\xp}{x_\text{part}}
\newcommand{\xr}{x_\text{reco}}

\newcommand{\qqquad}{\qquad\quad}

\def\d{\mathrm{d}}

\newcommand\one{\leavevmode\hbox{\small1\normalsize\kern-.33em1}}

\newcommand{\lag}{\mathscr{L}}

\newcommand{\loss}{\mathcal{L}} 	
\newcommand{\normal}{\mathcal{N}} 	

\newcommand{\arXiv}[2][]{%
	\ifthenelse{\equal{#1}{}}%
	{\href{http://arxiv.org/abs/#2}{arXiv:#2}}%
	{\href{http://arxiv.org/abs/#2}{arXiv:#2~[#1]}}}

\def\slashchar#1{\setbox0=\hbox{$#1$}           
   \dimen0=\wd0                                 
   \setbox1=\hbox{/} \dimen1=\wd1               
   \ifdim\dimen0>\dimen1                        
      \rlap{\hbox to \dimen0{\hfil/\hfil}}      
      #1                                        
   \else                                        
      \rlap{\hbox to \dimen1{\hfil$#1$\hfil}}   
      /                                         
   \fi}

\newcommand{\tikznode}[2]{%
\ifmmode%
\tikz[remember picture,baseline=(#1.base),inner sep=0pt] \node (#1) {$#2$};%
\else
\tikz[remember picture,baseline=(#1.base),inner sep=0pt] \node (#1) {#2};%
\fi}

\def\mathswitchr#1{\relax\ifmmode{\mathrm{#1}}\else$\mathrm{#1}$\xspace\fi}
\def\mathswitch#1{\relax\ifmmode#1\else$#1$\xspace\fi}

\graphicspath{{./figs/}}

\begin{document}


\begin{center}{\Large \textbf{
Simulation-Prior Independent Neural Unfolding Procedure
}}\end{center}

\begin{center}
Anja Butter\textsuperscript{1,2},
Theo Heimel\textsuperscript{3},
Nathan Huetsch\textsuperscript{1},
Michael Kagan\textsuperscript{4}, and
Tilman Plehn\textsuperscript{1,5}
\end{center}

\begin{center}
{\bf 1} Institut für Theoretische Physik, Universität Heidelberg, Germany
\\
{\bf 2} LPNHE, Sorbonne Universit\'e, Universit\'e Paris Cit\'e, CNRS/IN2P3, Paris, France \\
{\bf 3} CP3, Universit\'e catholique de Louvain, Louvain-la-Neuve, Belgium \\
{\bf 4} SLAC National Accelerator Laboratory, Menlo Park, CA, USA\\
{\bf 5} Interdisciplinary Center for Scientific Computing (IWR), Universit\"at Heidelberg, Germany
\end{center}

\begin{center}
\today
\end{center}


\section*{Abstract}
{\bf Machine learning allows unfolding high-dimensional spaces without binning at the LHC. The new SPINUP method extracts the unfolded distribution based on a neural network encoding the forward mapping, making it independent of the prior from the simulated training data. It is made efficient through neural importance sampling, and ensembling can be used to estimate the effect of information loss in the forward process. We showcase SPINUP for unfolding detector effects on jet substructure observables and for unfolding to parton level of associated Higgs and single-top production.}

\vspace{10pt}
\noindent\rule{\textwidth}{1pt}
\tableofcontents\thispagestyle{fancy}
\noindent\rule{\textwidth}{1pt}
\vspace{10pt}

\section{Introduction}
\label{sec:intro}

The vast data volumes and high-precision simulations at the LHC offer a unique opportunity to address some of the most fundamental questions in particle physics. At the same time, they present a challenge to our capacity for data analysis.
Monte Carlo generators form a robust foundation for simulation-based inference, modeling the full range of physics effects, from the hard scattering through parton showering and hadronization, all the way to the detector response~\cite{Campbell:2022qmc,Butter:2022rso}. However, the overall computational cost of these simulations is already high, and projected to grow beyond available computational resources with the upcoming HL-LHC. 

This motivates an alternative analysis strategy, based on data representations removing the detector or other, low-energy aspects of the forward simulation: (i) once unfolded, data can be used to test any number of new physics model very economically; (ii) unfolded data can be used by a broad physics community, not just inside the established collaborations. As a side effect, unfolded data from different experiments can be combined for instance for global SMEFT analyses~\cite{Brivio:2019ius, Elmer:2023wtr}, to answer the question to what level LHC data as a whole is described by the Standard Model.

Unfolding traditionally relied on matrix based approaches~\cite{Cowan:2002in, Spano:2013nca}.
While providing precise and accurate results when unfolding data represented as one dimensional histograms, such binned approaches do not work for many dimensions. Unfolding methods based on machine learning~\cite{Arratia:2021otl, Butter:2022rso} open up new ways to perform unbinned and multi-dimensional cross section measurements~\cite{Arratia:2021otl}. These unfolding methods mainly rely either on classifier based approaches~\cite{Andreassen:2019cjw} or generative neural networks~\cite{Bellagente:2019uyp}.
Classifiers can reweight a Monte Carlo simulation to the true underlying distribution~\cite{Andreassen:2019cjw, Andreassen:2021zzk}. Generative network can either learn a direct mapping between distributions at reco and truth level~\cite{Datta:2018mwd, Howard:2021pos, Diefenbacher:2023wec, Butter:2023ira, Butter:2024vbx} or learn a probability distribution at truth level for each reco level event~\cite{Bellagente:2019uyp, Bellagente:2020piv, Vandegar:2020yvw,
Backes:2022sph, Leigh:2022lpn, Ackerschott:2023nax, Shmakov:2023kjj, Shmakov:2024gkd}.
A comparison of different generative approaches can be found in Ref.~\cite{Huetsch:2024quz}.

While ML-unfolding lifts the restriction to low dimensional methods, some challenges of matrix based unfolding remain, in particular the impact of the simulation bias on unfolded data. Traditional unfolding techniques mitigate this prior dependence through Iterative Bayesian Unfolding~\cite{Lucy:1974yx, DAgostini:1994fjx}. Alternative approaches regularize the inversion matrix using singular value decomposition~\cite{Hocker:1995kb} or Tikhonov regularization~\cite{Schmitt:2012kp}. ML-approaches adapt iterative approaches~\cite{Andreassen:2019cjw, Backes:2022sph, Backes:2023ixi},  as well as bias reducing data processing techniques~\cite{Favaro:2025psi} which have been shown to work in experimental settings~\cite{H1:2021wkz, H1:2023fzk, LHCb:2022rky, Komiske:2022vxg, Song:2023sxb, Favaro:2025psi}. The disadvantage of iterative methods is that they are not data-efficient.

We explore a one-step approach to avoiding this bias in our Simulation-Prior Independent Neural Unfolding Procedure (SPINUP). It builds on the Neural Empirical Bayes (NEB)~\cite{Vandegar:2020yvw} approach and encodes the unfolded distribution as a generative model that can be forward-mapped to match the observed data. The unfolded distribution is then found by minimizing the difference between the folded distribution at the reconstruction level and the observed data. We develop a new ML-approach including importance sampling and pretraining. It alleviates the high computational cost associated with integrating the likelihood in every training step, and allows the method to scale to more challenging problems. We also show how ensembling tracks the unfolding uncertainty from the loss of information in the forward process. 

In this paper we first present the method and our extensions in Sec.~\ref{sec:methodology}. We then use a Gaussian toy example to showcase our method and the effect of ensembling in Sec.~\ref{sec:toys}. In Sec.~\ref{sec:omnifold} we apply SPINUP to the Omnifold jet substructure dataset~\cite{Andreassen:2019cjw}, an established unfolding benchmark. Finally, as a challenging physics example we unfold the full phase space of $tHj$ events to parton level. Here, we demonstrate how SPINUP recovers the parton-level distributon to high precision, without showing any prior dependence, and allows to measure the CP-phase in the top Yukawa coupling from unfolded data~\cite{Buckley:2015vsa,Ren:2019xhp,Bortolato:2020zcg,Bahl:2020wee,Martini:2021uey,Goncalves:2021dcu,Barman:2021yfh,Bahl:2021dnc,Kraus:2019myc}.

\section{Forward unfolding}
\label{sec:methodology}

We can understand unfolding using four probabilities, the simulated
part-level distribution\footnote{We use the term part-level to cover
  unfolding to particle level and to parton level.}, the simulated
reco-level distribution, the measured reco-level distribution in
data, and the unfolded part-level distribution. They are related
as~\cite{Plehn:2022ftl}
\begin{alignat}{9}
  & \psim(\xp)
  \quad \xleftrightarrow{\text{unfolding inference}} \quad 
  && \punf(\xp)
  \notag \\
  & \hspace*{-9mm} {\scriptstyle p(\xr|\xp)} \Bigg\downarrow
  && \hspace*{-5mm} {\scriptstyle \text{\; unfold}} \Bigg\uparrow
  \notag \\
  & \psim(\xr) 
  \quad \xleftrightarrow{\text{\; forward inference \;}} \quad 
  && \pd(\xr)
\label{eq:schematic}
\end{alignat}
The density $p(\xr|\xp)$ describes the forward mapping from the part-level phase space to the reco-level phase space. It is encoded implicitly in simulators in the form of 
paired events in $\xp$ and $\xr$. Like all unfolding methods, we assume that $p(\xr|\xp)$ is correctly described by the simulators. \\
Phrasing the problem in terms of our forward unfolding method, we want
to find a distribution in $\xp$ that reproduces $\pd(\xr)$ once it is
folded with the forward mapping.  Typically, there is no unique unfolding
solution, because limited statistics and information loss in the
forward mapping can lead to a range of valid $\punf(\xp)$ solutions.

\subsection{Neural Empirical Bayes}

We can solve this unfolding problem in a forward fashion, building on
the Neural Empirical Bayes (NEB)
approach~\cite{Vandegar:2020yvw}. Here, we encode the forward mapping
$p(\xr|\xp)$ and the unfolded distribution $\punf(\xp)$ in generative
neural networks and train the latter to reproduce $\pd(\xr)$.

\paragraph{Unfolding network}
The central object in forward unfolding is the generative network
\begin{align}
 p_{\theta}(\xp) \approx \punf(\xp)  \; ,
\end{align}
for which we adapt the parameters until the resulting reco-level
distribution matches the data,
\begin{align}
   p_{\theta}(\xr)  \; = \int \d \xp \; p(\xr|\xp) \;  p_{\theta}(\xp) \;   \stackrel{!}{=}  \; \pd(\xr) \; .
    \label{eq:forward_marginal}
\end{align}
For a valid solution, the folded distribution has to agree with the data. The loss
quantifies the discrepancy between these two reco-level distributions,
\begin{align}
  \kl [\pd(\xr),p_{\theta}(\xr)] 
  &= \underbrace{- \int \d \xr \: \pd(\xr) \; \log p_{\theta}(\xr)}_{\equiv \loss} + \text{const.}
    \label{eq:kl_loss}
\end{align}
where 
\begin{align}
  \log p_{\theta}(\xr) =
  \log \int \d \xp \;  p_\theta(\xp) \; p(\xr|\xp) \; .
    \label{eq:int_evi}
\end{align}
Minimizing the KL-divergence with respect to $\theta$ is equivalent to
maximizing the log-likelihood of the observed data distribution
$\pd(\xr)$ under our unfolding model $p_\theta(\xp)$.

\paragraph{Transfer network}
The main challenge of forward unfolding is the calculation of
$p_{\theta}(\xr)$ in Eq.\eqref{eq:int_evi} using the forward mapping
$p(\xr|\xp)$. Standard simulators encode this density implicitly. To access it explicitly, we
use paired simulated data $(\xp, \xr) \sim \psim(\xp, \xr)$ to train a
conditional generative neural network~\cite{Vandegar:2020yvw,
  Butter:2022vkj,Heimel:2023mvw} ahead of time,
\begin{align}
  p_\varphi(\xr|\xp) \approx p(\xr|\xp) \; .
    \label{eq:transfer_surrogate}
\end{align}
During the training of the unfolding network $p_\theta(\xp)$, the
transfer network $p_\varphi(\xr|\xp)$ is frozen. In the following we
will drop the subscript $\varphi$.

\paragraph{Monte Carlo integration}
The loss evaluation in Eq.\eqref{eq:kl_loss} uses
mini-batches. Following Eq.\eqref{eq:int_evi} we rely on Monte Carlo
integration, for each sample $\xr \sim \pd(\xr)$, to calculate
\begin{align}
  \log p_{\theta}(\xr) 
  = &\log
  \int \d \xp \;  p_\theta(\xp) \; p(\xr|\xp) \notag \\
%
%
 = & \lim_{N_\text{MC}\rightarrow \infty} \log \frac{1}{N_\text{MC}}
  \sum_{\{\xp \}}^{N_\text{MC}} \; p(\xr|\xp ) \; \Bigg|_{\xp \sim p_\theta(\xp)}  \; .
    \label{eq:MC_naive}
\end{align}
The algorithm samples a batch of points from $p_\theta(\xp)$ and
passes them through the transfer network to estimate $\log
p(\xr|\xp)$. These are aggregated using the numerically stable
logsumexp operation
\begin{align}
  \log p_{\theta}(\xr)
  = \text{logsumexp} \; \Big( p(\xr|x_\text{part, 1})\; , \;  ... \;, \;  p(\xr|x_\text{part, $N_\text{MC}$}) \Big) - \log N_\text{MC} \; .
 \label{eqq:MC_logsumexp}
\end{align}
Since the integration is inside of the $\log$, the MC approximation of
the integral is also inside the $\log$. This results in a biased but
consistent estimator for the marginal log likelihood, with the bias
rapidly decreasing towards zero as $N_\text{MC}$
increases~\cite{Vandegar:2020yvw}.

\subsection{SPINUP}

Our Simulation-Prior Independent Neural Unfolding Procedure (SPINUP)
uses state-of-the-art generative ML for the forward unfolding
described above. Challenges arise when the transfer function
$p(\xr|\xp)$ has narrow peaks. This increases the variance in the
Monte Carlo samples and renders a poor integral estimate. Increasing
the number of Monte Carlo samples is not a scalable solution, because
evaluating the loss from Eq.\eqref{eq:MC_naive} on a batch of reco-level
events involves calculating the transfer probability for each
reco-part pairing, so the memory cost is proportional to
$N_\text{Batch} \times N_\text{MC}$.

\paragraph{Neural importance sampling}
To improve the integration we employ neural importance sampling
(NIS)~\cite{Heimel:2022wyj, Heimel:2023ngj, Heimel:2024wph}, similar
to Refs.~\cite{Butter:2022vkj, Heimel:2023mvw}. It encodes a suitable
importance sampling $q(\xp|\xr)$ in a generative network to compute
the integral in Eq.\eqref{eq:int_evi}
\begin{align}
  \log p_{\theta}(\xr) = \log
  \int \d \xp \; q(\xp|\xr) \; \frac{p_\theta(\xp) \; p(\xr|\xp)}{q(\xp|\xr)} \; .
\label{eq:IS_integral}
\end{align}
Choosing the sampling distribution such that the integrand is constant
minimizes the variance of the integral. The ideal sampling
distribution should therefore be proportional to the numerator
\begin{align}
 q(\xp|\xr) 
  &\propto p_\theta(\xp) \; p(\xr|\xp) \notag \\
  &\propto p_\theta(\xp|\xr) \; .
  \label{eq:ideal_IS}
\end{align}
In the second step we have applied Bayes' theorem to eliminate the direct dependence on $p_\theta(\xp)$, but the conditional distribution, $p_\theta(\xp|\xr)$ is still defined
implicitly via $p_\theta(\xp)$ and $p(\xr|\xp)$ and changes during the
training of the unfolding network. For the NIS network we use the
paired simulations $(\xp, \xr)$ to train it on the inverse conditional
distribution~\cite{Bellagente:2019uyp,Bellagente:2020piv} 
\begin{align}
  q_\psi(\xp|\xr) \approx \psim(\xp|\xr) \; .
    \label{eq:IS_surrogate}
\end{align}
It is proportional to $\psim(\xp)$ instead of
$p_\theta(\xp)$. Therefore, it will not give the ideal sampling
distribution, but as long as $p_\theta(\xp)$ is not too far from its
simulation counterpart, it will serve as an efficient importance
sampling distribution. Unlike for generative unfolding, this
conditional generative network serves as a numerical tool for
training $p_\theta(\xp)$. Any prior dependence will decrease the
training efficiency, but not affect the unfolding result.

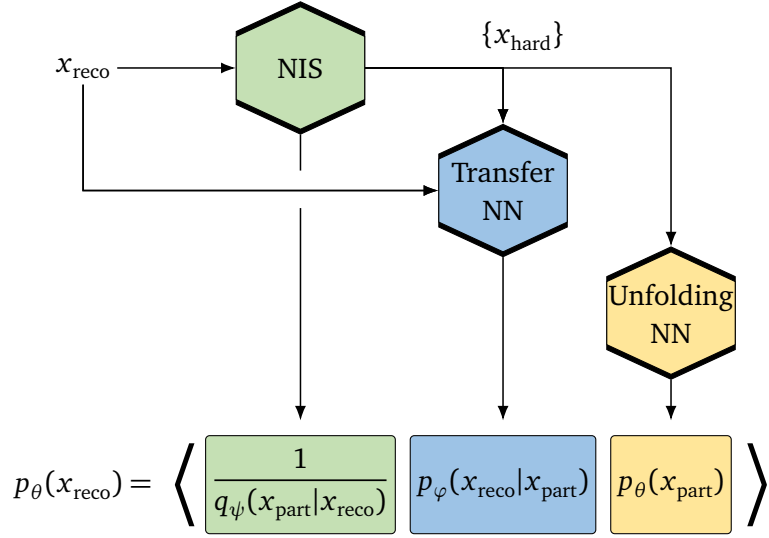
\begin{figure}[t]
    \centering
    \definecolor{Rcolor}{HTML}{E99595}
\definecolor{Gcolor}{HTML}{C5E0B4}
\definecolor{Bcolor}{HTML}{9DC3E6}
\definecolor{Ycolor}{HTML}{FFE699}

\tikzstyle{expr} = [rectangle, rounded corners=0.3ex, minimum width=2.2cm, minimum height=2cm, text centered, align=center, inner sep=0, fill=Ycolor, font=\LARGE, draw]
\tikzstyle{txt_huge} = [align=center, font=\Huge, scale=2]
\tikzstyle{txt} = [align=center, font=\LARGE]
\tikzstyle{cinn} = [double arrow, double arrow head extend=0cm, double arrow tip angle=130, shape border rotate=90, inner sep=0, align=center, minimum width=2.6cm, minimum height=2.5cm, fill=Gcolor, draw, font=\LARGE]
\tikzstyle{cinn_black} = [cinn, minimum height=2.7cm, fill=black]
\tikzstyle{arrow} = [thick,-{Latex[scale=1.0]}, line width=0.2mm, color=black]
\tikzstyle{line} = [thick, line width=0.2mm, color=black]

\begin{tikzpicture}[node distance=2cm, scale=0.65, every node/.style={transform shape}]

\node (likeli) [txt] {$p_{\theta}(x_\text{reco}) = $};
\node (langle) [txt_huge, right of=likeli, xshift = -1.0cm] {$\langle$};
\node (proposal_fct) [expr, right of=langle, xshift =0.4cm, fill=Gcolor ] {\LARGE $\;\dfrac{1}{q_\psi(x_\text{part}|x_\text{reco})}\;$};
\node (transfer_fct) [expr, right of=proposal_fct, xshift = 2.1cm, fill=Bcolor] {\LARGE $\;p_\varphi(x_\text{reco}|x_\text{part})\;$};
\node (unfolded) [expr, right of=transfer_fct, xshift = 1.4cm, fill=Ycolor] {\LARGE $\;p_\theta(x_\text{part})\;$};
\node (langle) [txt_huge, right of=unfolded, xshift = -1.1cm] {$\rangle$};

\node (unfcinn_b) [cinn_black, above of=unfolded, yshift=1.5cm] {};
\node (unfcinn) [cinn, above of=unfolded, yshift=1.5cm, fill=Ycolor] {Unfolding\\NN};
\node (transfer_b) [cinn_black, above of=transfer_fct, yshift=4cm] {};
\node (transfer) [cinn, above of=transfer_fct, yshift=4cm, fill=Bcolor] {Transfer\\NN};
\node (nis_b) [cinn_black, above of=proposal_fct, yshift=6.5cm] {};
\node (nis) [cinn, above of=proposal_fct, yshift=6.5cm, fill=Gcolor] {NIS};

\node (xr) [txt, above of=likeli, yshift=6.5cm] {$x_\text{reco}$};

\draw [arrow, color=black] (xr.east) -- (nis.west);
\draw [arrow, color=black] (xr.south) -- (xr.south |- transfer_b.west) -- (transfer_b.west);
\draw [arrow, color=black] (xr.south) -- (xr.south |- transfer_b.west) -- (transfer_b.west);
\draw [arrow, color=black] (nis.east) -- (transfer_b.north |- nis.east) -- (transfer_b.north);
\draw [arrow, color=black] (nis.east) -- (transfer_b.north |- nis.east) -- (transfer_b.north)node[midway,above,xshift=0.4cm, yshift=0.8cm]{\LARGE $\{x_\text{hard}\}$};
\draw [arrow, color=black] (nis.east) -- (unfcinn_b.north |- nis.east) -- (unfcinn_b.north);
\draw[arrow, color=black](transfer_b.south) -- ([yshift=0.3cm]transfer_fct.north);
\draw[arrow, color=black](unfcinn_b.south) -- ([yshift=0.3cm]unfolded.north);
\draw[arrow, color=black]([yshift=-1.5cm]nis_b.south) -- ([yshift=0.3cm]proposal_fct.north);
\draw [line, color=black] (nis_b.south) -- ([yshift=5.4cm]proposal_fct.north);

\end{tikzpicture}
    \caption{Reco-level maximum likelihood training of the unfolding
      network $p_\theta(\xp)$.}
    \label{fig:pipeline}
\end{figure}

Combining Eq.\eqref{eq:IS_integral} with neural importance sampling,
the training of the forward unfolding of a single reco-level event now
minimizes
\begin{align}
\loss_\text{SPINUP} 
  = &- \int \d \xr \: \pd(\xr) \log p_{\theta}(\xr)  \notag \\
    \label{eq:reco_MC_IS}
  = &- \int \d \xr \: \pd(\xr) \log  \int \d \xp \; q_\psi(\xp|\xr)
  \frac{p(\xr|\xp)p_\theta(\xp)}{q_\psi(\xp|\xr)} \\
  = & - \lim_{N_\text{MC}\rightarrow \infty}  \int \d \xr \: \pd(\xr) \log  \frac{1}{N_\text{MC}}\sum_{\{ \xp \}}^{N_\text{MC}} \; \frac{p(\xr|\xp)p_\theta(\xp)}{q_\psi(\xp|\xr)} \; \Bigg|_{\xp \sim q_\psi(\xp|\xr)} \; . \notag 
\end{align}
The training is illustrated in Fig.~\ref{fig:pipeline}. The NIS
network $q_\psi(\xp|\xr)$ and the transfer network $p_\varphi(\xr |
\xp)$ are trained ahead of time, and their parameters are frozen
during the main training of the unfolding network $p_\theta(\xp)$.

\paragraph{Pre-training}
While NIS significantly eases its computational cost, the loss
calculation in Eq.\eqref{eq:reco_MC_IS} remains a computational
challenge, and convergence is slow. In Eq.\eqref{eq:MC_naive} we 
have introduced a sum over $N_\text{MC}$ samples to 
evaluate the $\xp$-integration. To speed up convergence, we 
pre-train the unfolding network on a single part-level event, 
\ie $N_\text{MC}=1$. The sum within the logarithm is then reduced to a single summand and the calculation of the gradient simplifies substantially
\begin{align}
\loss_\text{Pre} 
  &= - \int \d \xr \: \pd(\xr) \; \log \frac{p(\xr|\xp)p_\theta(\xp)}{q_\psi(\xp|\xr)} \; \Bigg|_{\xp \sim q_\psi(\xp|\xr)} \nonumber\\
\Rightarrow \quad
  \nabla_\theta \; \loss_\text{Pre} &=   - \int \d \xr \: \pd(\xr) \;\nabla_\theta \log
  p_\theta(\xp) \; \Bigg|_{\xp \sim q_\psi(\xp|\xr)} \; .
\label{eq:MC1_IS_gradient}
\end{align}
For the gradient of the pre-training loss, the two conditional
probabilities do not contribute. The setup reduces to the likelihood
training of $p_\theta(\xp)$ on the part-level distribution obtained by
unfolding $\pd(\xr)$ with $q_\psi(\xp|\xr)$. After pre-training,
$p_\theta(\xp)$ corresponds to the outcome of generative unfolding
acting on $\pd(\xr)$, including a potential prior dependence. For
SPINUP it serves as a network initialization only. To validate the prior independence given numerical limitations of network trainings we have cross-checked that the pre-training only impacts the convergence time and not the result itself.

\paragraph{Categorical approximation}
To further improve the ML-computation of the loss we employ another
approximation~\cite{burda2016importanceweightedautoencoders}. The
log-likelihood for one reco-event in Eq.\eqref{eq:reco_MC_IS} reads
\begin{align}
   \loss_\text{SPINUP} \approx \log \frac{1}{N_\text{MC}} \sum_{i}^{N_\text{MC}} w_i 
   \qquad \text{with} \qquad  w_i = \frac{p(\xr | \xp ) \;  p_\theta (\xp)}{q_\psi(\xp| \xr)} \Bigg |_{\xp \sim q_\psi(\xp| \xr)}
\end{align}
We can rewrite its gradient as
\begin{align}
   \nabla_\theta \loss_\text{SPINUP} 
   \approx
   \frac{1}{\sum_{i}^{N_{\text{MC}}} w_i }  \sum_{i}^{N_\text{MC}} \nabla_\theta w_i 
   =  \sum_{i}^{N_\text{MC}} \frac{w_i}{\sum_{i}^{N_\text{MC}} w_i }  \nabla_\theta \log w_i 
\end{align}
The gradient $\nabla_\theta \log p_\theta$ is given by the
expectation value of $\nabla_\theta \log w_i $ under a discrete
categorical distribution with $N_\text{MC}$ classes with class
probabilities $w_i/\sum_i^{N_\text{MC}} w_i$. We can approximate it
with a single draw from the distribution as
\begin{align}
   \nabla_\theta \loss_\text{SPINUP} 
   \approx \nabla_\theta \log w_i 
   \qquad \text{with} \qquad w_i \sim \mathcal{C} \left( \frac{w_1}{\Bar{w}}, ..., \frac{w_{MC}}{\Bar{w}} \right)
   \label{eq:categorical}
\end{align}
For this unbiased estimator of the original Monte Carlo estimator
gradient we still draw $N_\text{MC}$ samples from the importance
sampling distribution to calculate the categorical weights in
Eq.\eqref{eq:categorical}, but we only back-propagate gradients with
respect to one of these samples. This approximation increases the
estimator variance, but saves memory and in turn allows for a larger
batch size or more Monte Carlo samples. We find this a favorable
trade-off.

Finally, we achieve a speed-up through multiple backpropagation steps
per batch, using the same categorical distribution. Since this does
not require drawing new samples from the importance sampling network
or re-evaluating the transfer density, it comes at very little
additional cost.

\paragraph{SPINUP}
The above steps define an algorithm, where we assume access
to paired simulated events at part- and reco-level as well as to data
at reco-level.
\begin{enumerate}
    \item Train the generative transfer network $p_\varphi(\xr|\xp)$ on
      the paired simulated data;
    \item Train the generative NIS network $q_\psi(\xp|\xr)$ on the
      paired simulated data;
    \item Pre-train the generative unfolding network $p_\theta(\xp)$ with 
      $N_\text{MC}=1$ as an initialization;
    \item Train the unfolding network $p_\theta(\xp)$ to reproduce the
      data, using the categorical approximation
       of the maximum likelihood loss, Eq.\eqref{eq:reco_MC_IS}.
\end{enumerate}
Notably, this will avoid a prior dependence without any iteration.

\paragraph{Network architectures}
SPINUP includes three separate generative networks, the transfer
network $p_\varphi(\xr|\xp) \approx p(\xr|\xp)$, the NIS network
$q_\psi(\xp|\xr) \approx \psim(\xp|\xr)$, and the unfolding network
$p_\theta(\xp)$. Our choice of architectures is guided by the specific
role played by these networks.

Diffusion networks are the state of the art in generative HEP
applications~\cite{Butter:2023fov}, but they are not suitable for any of the networks in our setup due to their
slow density evaluation. Here, normalizing flows are ideal, achieving comparable precision at an
order of magnitude reduced computational cost~\cite{Heimel:2023mvw}.

For the transfer and unfolding networks, our loss calculation requires
fast and precise density estimation, but does not require sample
generation. We employ the Transfermer network~\cite{Heimel:2023mvw},
which has shown an excellent precision-speed trade-off in an
application closely related to ours. It combines a Transformer
backbone with one-dimensional RQS-splines, that autoregressively
encode the density. The slow sampling coming with autoregressive
generation does not matter for our application, since these networks
are only used for density estimation.

The NIS network generates large numbers of samples in each training
iteration. It requires an architecture for fast generation,
potentially trading-off some precision. We employ an INN with RQS
spline blocks~\cite{Heimel:2022wyj, Heimel:2023ngj, Heimel:2024wph}.

\paragraph{Ensembling}
We employ ensembling for a rough estimate of the uncertainty
introduced by our methodology. Typically, ensembling is used to estimate the uncertainty associated with the stochastic network
initialization and training. \\ In our setup, it gives a lower bound on one additional source
of uncertainty, the loss of information in the forward mapping. Since
the loss calculation happens purely at the reco-level, any unfolded
distribution $p_\theta(\xp)$ which reproduces the observed data
minimizes the loss. If the ensemble is well-trained, every network
will reproduce the same reco-level distribution, while mapping out a
space of possible solutions at part-level. Such an ensemble comes with
no guarantee to find every possible part-level solution, so the
ensemble uncertainty should be viewed as a lower bound on the true
information loss uncertainty.

\section{Gaussian illustration}
\label{sec:toys}

\begin{figure}[!b]
    \includegraphics[width=0.325\textwidth]{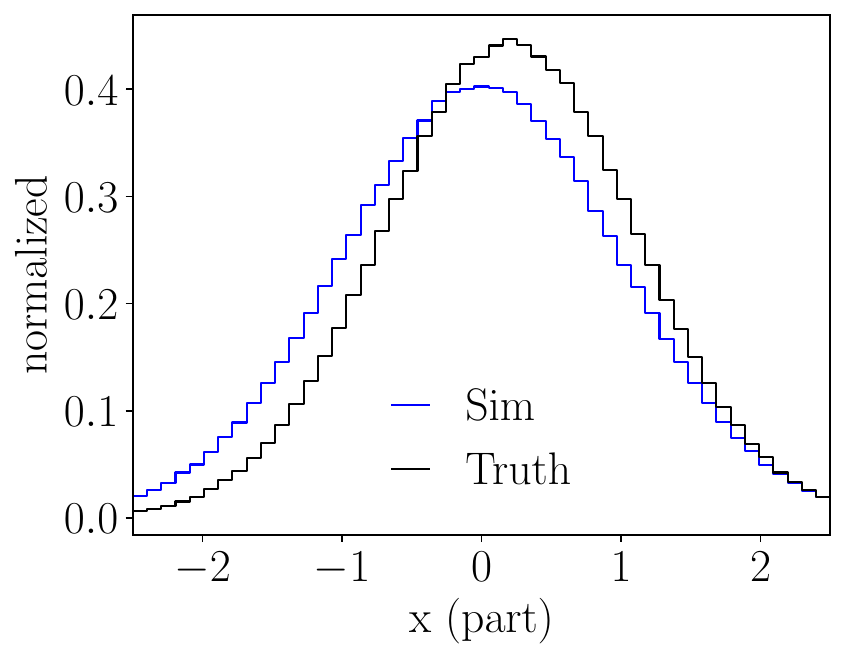}
    \includegraphics[width=0.325\textwidth, page=1]{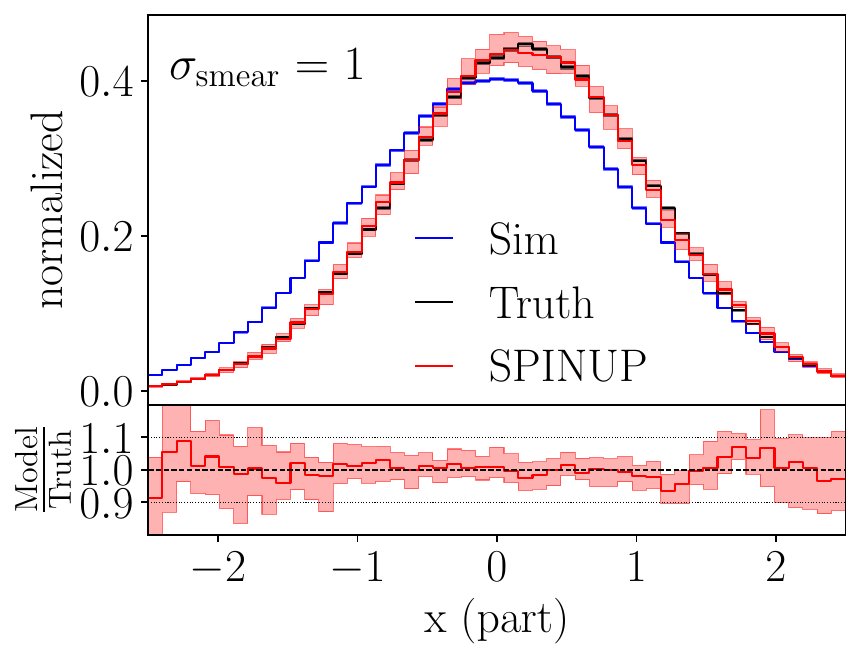}
    \includegraphics[width=0.325\textwidth, page=2]{figs/toy_gauss/observables_gausstoy.pdf} \\
    \includegraphics[width=0.325\textwidth]{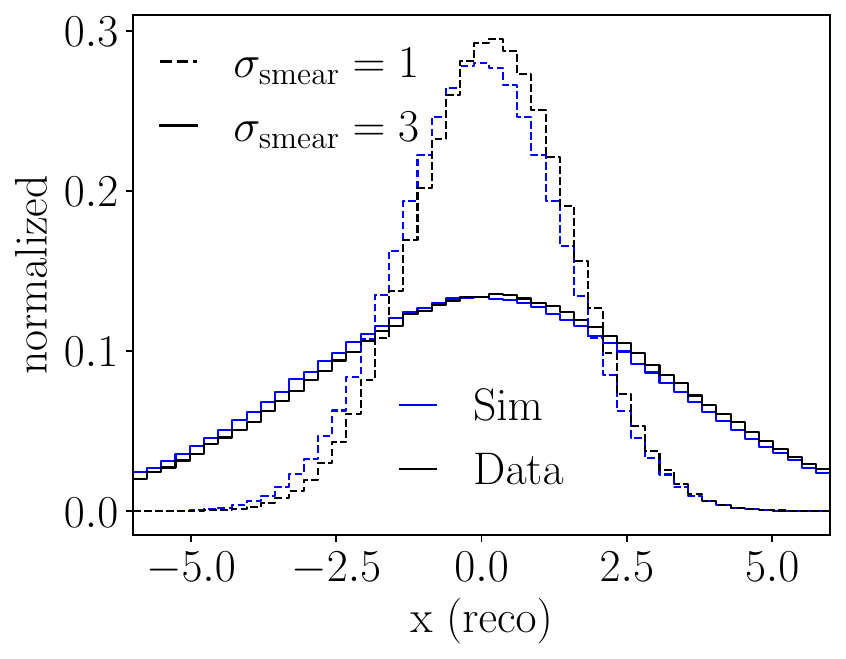}
    \includegraphics[width=0.325\textwidth, page=3]{figs/toy_gauss/observables_gausstoy.pdf}
    \includegraphics[width=0.325\textwidth, page=4]{figs/toy_gauss/observables_gausstoy.pdf} 
    \caption{Unfolding results for the Gaussian toy for
      $\sigma_\text{smear}=1,3$. The upper row shows the part-level
      unfolded distributions, the lower row the obtained reco-level
      distributions. Uncertainties are obtained from an ensemble of 32
      networks.}
    \label{fig:gauss_toy_results}
\end{figure}

An instructive example for forward unfolding is a Gaussian toy
model. We assume that the simulated and target distributions are
normal distributions,
\begin{align}
   \psim(\xp) &= \normal(\xp ; 0, 1) \notag  \\ 
   p_\text{truth}(\xp) &= \normal(\xp ; 0.2, 0.9) \; .
   \label{eq:gauss_toy_part}
\end{align}
The difference represents a small discrepancy between simulation and
truth. Also the detector effects are assumed to be a simple Gaussian
smearing,
\begin{align}
   p(\xr|\xp) = \normal(\xr ; \xp, \sigma_\text{smear}) \; .
   \label{eq:gauss_toy_transfer}
\end{align}
We can then calculate the reco-level distributions as
\begin{align}
   \psim(\xr) 
   &= \int \d \xp \; \psim(\xp) \; p(\xr|\xp) 
   = \normal \left(\xr ; 0, \sqrt{1 + \sigma_\text{smear}^2}\right) \notag\\
   \pd(\xr) 
   &= \int \d \xp \; p_\text{truth} (\xp) \; p(\xr|\xp) 
   = \normal \left(\xr ; 0.2, \sqrt{0.9^2 + \sigma_\text{smear}^2} \right) \; .
   \label{eq:gauss_toy_reco}
\end{align}
During unfolding we observe the two reco-level distributions in
Eq.\eqref{eq:gauss_toy_reco} and know the transfer function
Eq.\eqref{eq:gauss_toy_transfer} from simulated paired events. The task is to
infer $p_\text{truth}(\xp)$, with the known outcome
\begin{align}
    \sigma_\text{truth}^2 = \sigma_\text{data}^2 - \sigma_\text{smear}^2 \; .
    \label{eq:sigma_diff}
\end{align}

The part-level distributions Eq.\eqref{eq:gauss_toy_part} and the
corresponding reco-level distributions Eq.\eqref{eq:gauss_toy_reco}
for $\sigma_\text{smear}=1$ and~3 are shown in the left panels of
Fig.~\ref{fig:gauss_toy_results}. At part-level there is a clear
difference between the simulated and truth distributions. For a
precise detector with $\sigma_\text{smear}=1$ the two distributions
are still clearly separated at reco-level. For
$\sigma_\text{smear}=3$, the two reco-level distributions are hard to
distinguish, implying that reconstructing the truth will be difficult.

In the center and the right column of Fig.~\ref{fig:gauss_toy_results}
we show unfolding results from an ensemble of 32 networks.
The upper row shows the unfolded, part-level distributions, the lower
row the corresponding reco-level distributions after convoluting with
the forward simulation. For both smearing strengths, the ensemble of
reco-level solutions matches the observed data without a significant
spread.  This implies that the unfolded distributions are valid.  At
part-level we see a large difference between the ensemble spreads for
the two smearing strengths. Ensembles come with no guarantee to map
out the space of possible solutions, so these uncertainties should be
taken with caution, but they confirm that a large loss of information
in the forward simulation causes a large uncertainty on the unfolded
distribution.\\
To see that the algorithm has solved its task, we fit Gaussians to the SPINUP solutions at both reco- and part-level. For the widths we find
\begin{alignat}{4}
    \sigma_\text{smear}=1:  \quad &
  \sigma^2_\text{reco} = 1.813 \pm 0.0169 \; (\text{truth } 1.81) &\quad  &\sigma^2_\text{part}= 0.814 \pm  0.0166 \; (\text{truth } 0.81) & \nonumber \\
  \sigma_\text{smear}=3:  \quad &
  \sigma^2_\text{reco} = 9.853 \pm 0.095 \; (\text{truth } 9.81)& \quad  &\sigma^2_\text{part} = 0.859 \pm 0.091 \; (\text{truth } 0.81) \; . \nonumber
    \label{eq:gaussian_inference}
\end{alignat}
At reco-level, for both smearing strengths SPINUP recovers the correct squared width with a relative precision of roughly 1 percent. Applying Gaussian error propagation to Eq.~\ref{eq:sigma_diff} yields
\begin{align}
    \Delta\sigma_\text{truth}^2  + \Delta\sigma_\text{smear}^2= \Delta \sigma_\text{data}^2  \; .
\end{align}
Assuming no error on the smearing widths, the absolute error on the squared widths at reco- and at part-level should be the same. We observe this to be the case for the SPINUP uncertainties. Since the reco-level widths are significantly larger than the part-level width, this propagation of the absolute value of the uncertainties results in a larger relative uncertainty at part-level. For the $\sigma_\text{smear}=1$ case we observe a relative uncertainty of roughly 2 percent at part-level, while for the $\sigma_\text{smear}=3$ case the relative uncertainty is about 10 percent. 

\paragraph{Information loss}
For a drastic information loss we consider a mixture of two Gaussians,
\begin{figure}[!t]
    \includegraphics[width=0.495\textwidth]{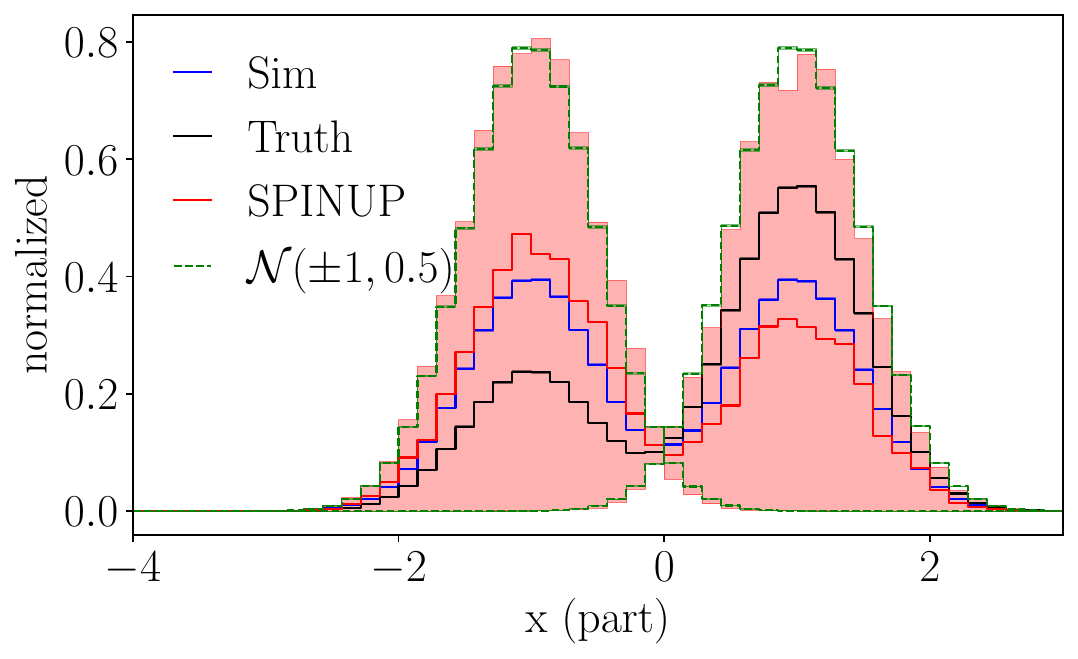}
    \includegraphics[width=0.495\textwidth]{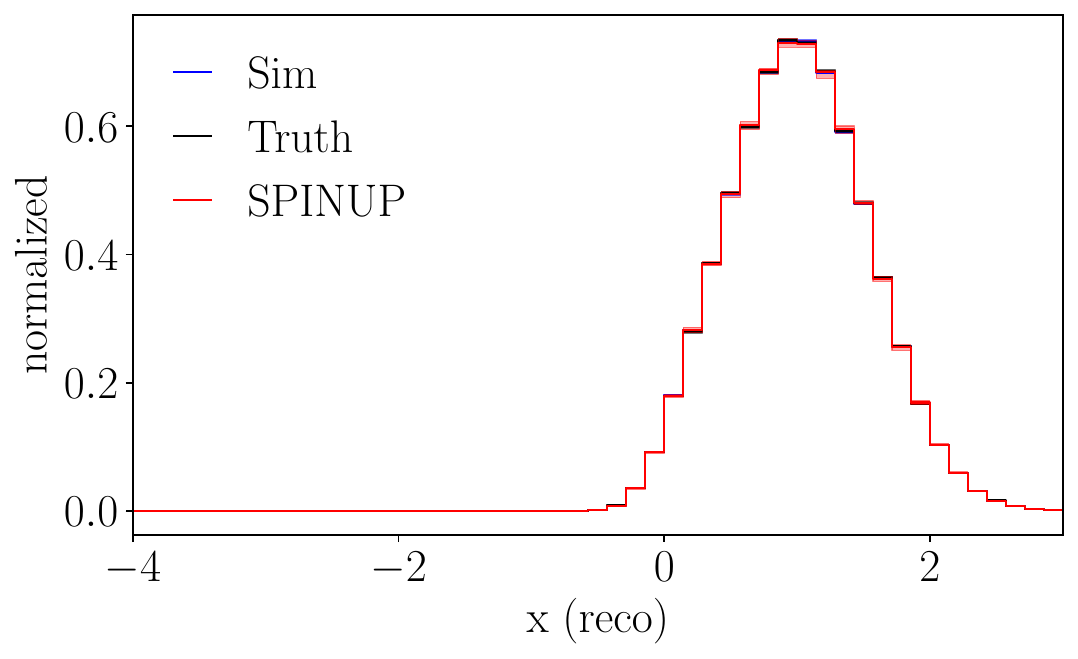}
    \caption{Part-level and reco-level distributions for the sign loss
      toy example Eqs.\eqref{eq:toy_bimodal_part}
      and~\eqref{eq:toy_bimodal_forward}. Unfolding results are shown
      for an ensemble of 32 networks.}
    \label{fig:sign_loss_toy}
\end{figure}
%
\begin{align}
  p(\xp|\alpha) &= \alpha \; \normal(\xp \; ; -1, 0.5) \; + \; (1-\alpha) \; \normal(\xp \; ; 1, 0.5) \notag \\
  & \qquad \text{with} \qquad 
  \alpha_\text{sim} = 0.5 \qqquad \alpha_\text{truth} = 0.3 \; .
    \label{eq:toy_bimodal_part}
\end{align}
We consider a forward mapping that loses the sign and adds Gaussian
smearing,
\begin{align}
  \xr = | \xp | + \normal(0,0.2) \; .
    \label{eq:toy_bimodal_forward}
\end{align}
By construction, every $\alpha$ will lead to the
same reco-level distribution.

The toy dataset and the SPINUP unfolding results are shown in
Fig.~\ref{fig:sign_loss_toy}. The left panel shows the part-level
distributions, the right panel the corresponding reco-level
distributions after convoluting with the forward simulation. At
part-level, the ensemble of 32 networks shows a huge spread,
indicating a wide range of valid solutions. It turns out that the
ensemble does map out the entire solution space. This is showcased by
the two green curves, the boundary cases $\mathcal{N}(\pm1, 0.5)$ for
$\alpha = 0,1$. The right plot confirms that all of these part-level
solutions collapse to the identical observed distribution at
reco-level.

\section{Jet substructure dataset}
\label{sec:omnifold}

As a physics toy example we look at the OmniFold
dataset~\cite{Andreassen:2019cjw}, a simple benchmark for unfolding
tools~\cite{Vandegar:2020yvw, Huetsch:2024quz, Diefenbacher:2023wec,
  Desai:2024yft, Butter:2024vbx}. It consists of 1.6M simulated LHC events for the process
\begin{align} 
 pp \to Z + \text{jets}
\end{align}
at 14~TeV.  To emulate the difference between Monte Carlo simulation
and data, we use two versions: first, events are generated with
Pythia~8.244~\cite{Sjostrand:2014zea} and Tune~26; second, events are
generated with Herwig~7.1.5~\cite{Bellm:2015jjp} and standard
tune. Both datasets use Delphes~3.5.0~\cite{deFavereau:2013fsa} with
the CMS card with particle-flow for the detector simulation. Jets are
clustered before and after Delphes using the anti-$k_T$
algorithm~\cite{Cacciari:2008gp} with $R= 0.4$, implemented in
FastJet~3.3.2~\cite{Cacciari:2011ma}.  We consider the Herwig events
at reco-level as data. The corresponding Herwig part-level events
represent the unknown truth. The paired Pythia events serve as the
simulation dataset,
\begin{align}
  p_\varphi(\xr|\xp) \approx p_\text{Pythia}(\xr|\xp) \; .
\end{align}

The goal is to unfold jet substructure measurements. The six
observables used are the jet mass, width, multiplicity, soft drop
mass~\cite{Larkoski:2014wba,Dasgupta:2013ihk}, momentum fraction, and
the $N$-subjettiness ratio~\cite{Thaler:2010tr},
\begin{align}
  \left\{ \quad
  m, \;
  \tau_{1}^{(\beta=1)},\;
  N, \;
  \rho = \frac{m_\text{SD}^2}{p_T^2} \;
  z_g, \; 
  \tau_{21} = \frac{\tau_2^{(\beta=1)}}{\tau_1^{(\beta=1)}} 
  \quad \right\} 
\end{align}
with $z_\text{cut}= 0.1$ and $\beta = 0$ for the momentum fraction.

\begin{figure}[!b]
    \includegraphics[width=0.33\textwidth, page=1]{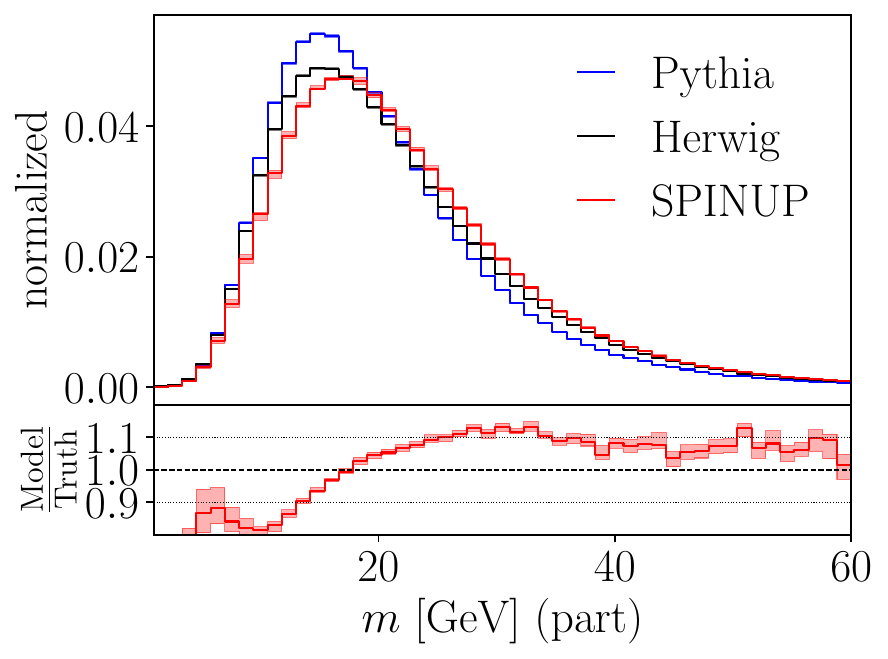}
    \includegraphics[width=0.33\textwidth, page=1]{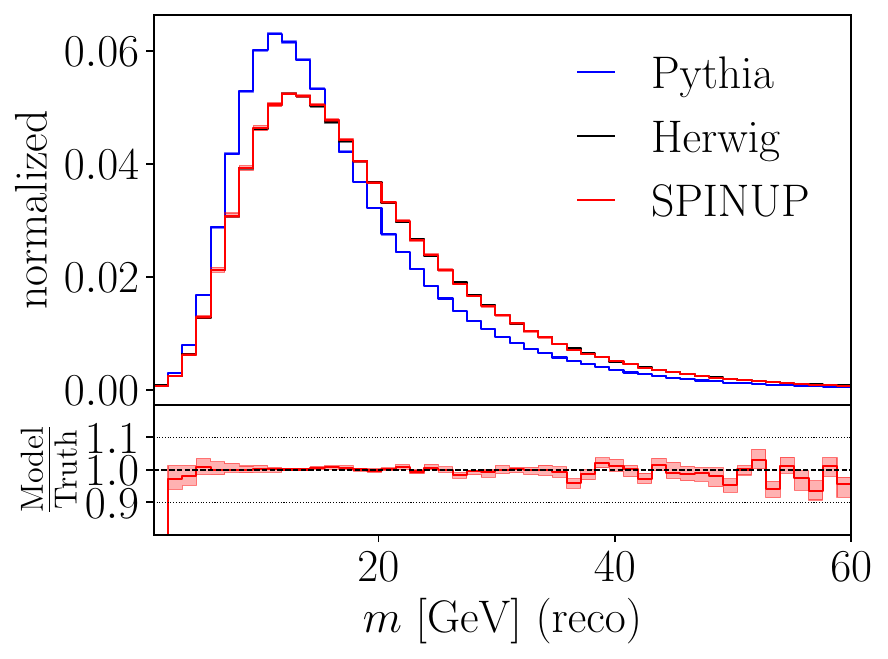}
    \includegraphics[width=0.33\textwidth, page=1]{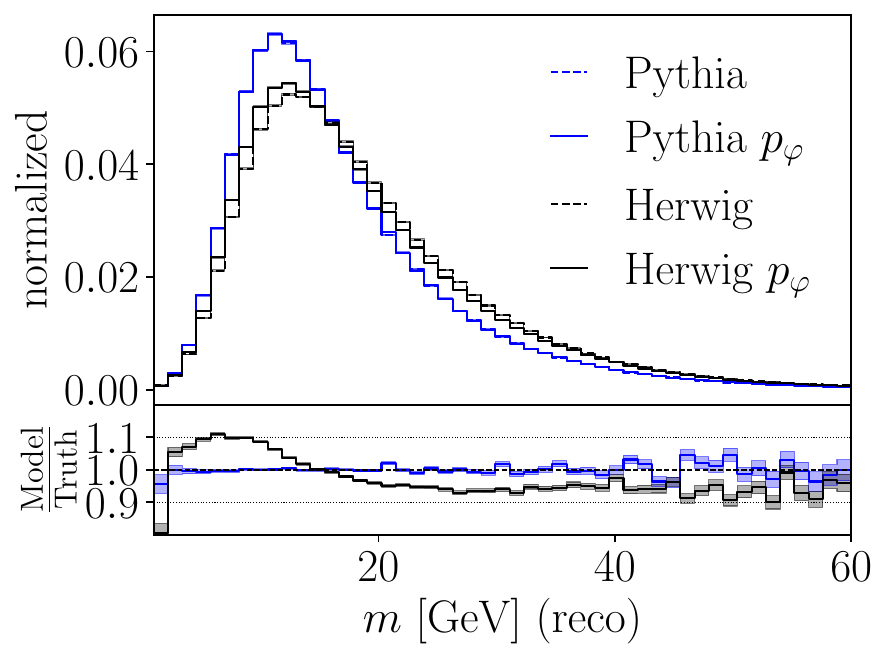}
    \caption{Left: part-level unfolding result for the jet
      mass. Center: reco-level distribution after passing the
      unfolding through the forward simulation. Right: transfer
      network $p_\varphi(\xr|\xp)$ trained on Pythia and evaluated on
      both Pythia and Herwig. The full set of observables is shown in
      Figs.~\ref{fig:omnifold_transfer_full}
      and~\ref{fig:omnifold_part_trueHerwig} in the Appendix.}
    \label{fig:omnifold_true_Herwig}
\end{figure}

We train SPINUP on this 6-dimensional phase space. The transfer and
NIS networks are pre-trained on the Pythia event pairs. The unfolding
training uses the Herwig reco-level events. The unfolded jet mass
distribution is shown in the left panel of
Fig.~\ref{fig:omnifold_true_Herwig}. We see that SPINUP does not recover the correct truth, with
deviations up to $10\%$.

To identify the underlying issue, we push the unfolded events through
the learned transfer function and look at the reco-level distribution in the
center panel of Fig.~\ref{fig:omnifold_true_Herwig}. We observe
perfect closure, so with the given forward simulation SPINUP has
fulfilled this task perfectly.

It could be that the solution space is degenerate, so SPINUP finds an
unfolded distribution different from the (Herwig) truth, but still
compatible with the data. However, in this particular dataset the
problem is the forward simulation, as shown in the right panel of
Fig.~\ref{fig:omnifold_true_Herwig}. It shows the distribution we find
when applying the trained forward surrogate $p_\varphi(\xr|\xp)$ to
the Pythia and Herwig part-level datasets. For Pythia, it reproduces
the reco-level distribution at the percent-level, but for Herwig we
observe a significant discrepancy. This means the forward conditional
probability does not generalize, violating a fundamental assumption of
all unfolding algorithms.

\paragraph{Forward simulation}
The reason that the forward simulation does not generalize is the
high-level summary statistics. In both datasets the detector effects
are simulated with the same Delphes tune, so in the full low-level
phase space the four-momenta of all individual particles are
consistent between Pythia and Herwig.  Once we look at jet
observables, the loss of information induces a prior dependence in the
forward simulation.

\begin{figure}[b!]
    \includegraphics[width=0.32\textwidth, page=1]{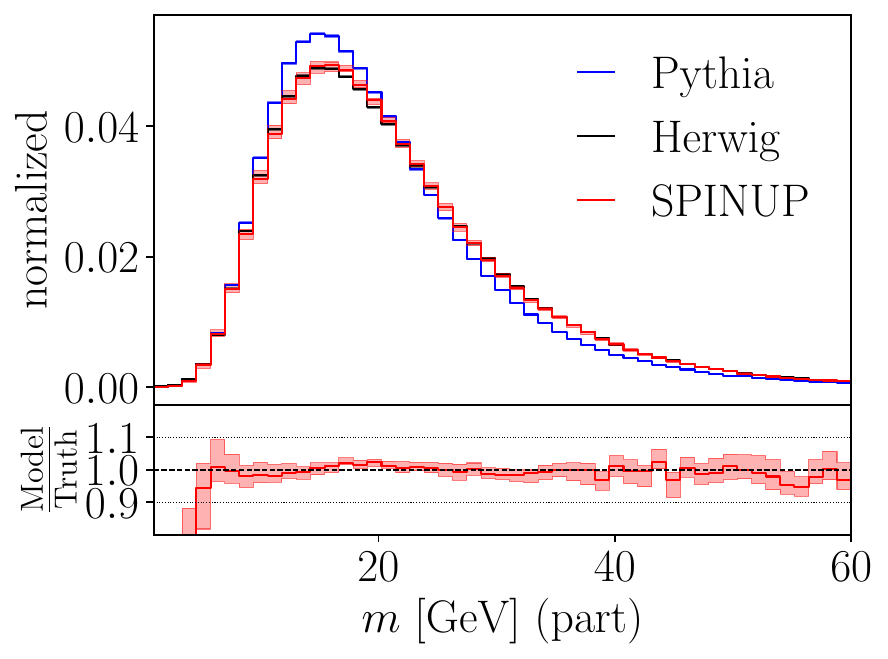}
    \includegraphics[width=0.32\textwidth, page=2]{figs/omnifold/observables_part_omnifold.pdf}
    \includegraphics[width=0.32\textwidth, page=3]{figs/omnifold/observables_part_omnifold.pdf}\\
    \includegraphics[width=0.32\textwidth, page=4]{figs/omnifold/observables_part_omnifold.pdf}
    \includegraphics[width=0.32\textwidth, page=5]{figs/omnifold/observables_part_omnifold.pdf}
    \includegraphics[width=0.32\textwidth, page=6]{figs/omnifold/observables_part_omnifold.pdf}
    \caption{Part-level unfolded distributions from the fake Herwig reco-distribution.}
    \label{fig:omnifold_part}
\end{figure}

This effect of a lossy summary statistics $s_\text{part}(\xp)$ can be
seen by including it as an explicit condition in the forward
simulation
\begin{align}
  p(\xr|s_\text{part})
  &= \int \d \xp \; p(\xr|\xp) \; p(\xp|s_\text{part}) \; .
  \label{eq:summary_stat}
\end{align}
%
The conditional probability $p(\xr|\xp)$ is the complete forward mapping,
while $p(\xp|s_\text{part})$ inverts the
computation of the summary statistics,
\begin{align}
  p(\xp|s_\text{part}) = p(s_\text{part}|\xp) \; \frac{p(\xp)}{p(s_\text{part})}  \; .
\end{align}
In case of information loss, $p(s_\text{part}|\xp)$ is well-defined, while 
the inverse includes degeneracies or flat directions. 
Without loss of information, this
relation reduces to a change of variables.
The prior dependence on $p(\xp)$ implies 
\begin{align}
  p_\text{Pythia}(\xp|s_\text{part}) &\neq p_\text{Herwig}(\xp|s_\text{part}) 
\notag \\
  \Rightarrow \qquad 
  p_\text{Pythia}(\xr|s_\text{part}) &\neq p_\text{Herwig}(\xr|s_\text{part}) \; .
\end{align}

This shortcoming could be addressed by unfolding the full phase space, leading to
the challenge of scaling generative networks to $\mathcal{O}(100)$
particles. To stick to our established benchmark, we instead circumvent 
this issue with a slight change to the
dataset: instead of using the reco-level Herwig data, we use the
trained 6-dimensional forward surrogate $p_\varphi(\xr|\xp) \approx
p_\text{Pythia}(\xr|\xp)$ to generate synthetic reco-level jet
substructure observables for the Herwig sample. This way, the forward
simulation generalizes.

The consistent SPINUP results are shown in
Fig.~\ref{fig:omnifold_part}. Again, we train an ensemble of 32 unfolding networks and show 
the median together with the centered 68\% range.
We recover the correct Herwig part-level
distributions over the entire phase space at the per-cent
level. For most observables, the fluctuations are covered
by the ensemble.

\section{Unfolding to parton level}
\label{sec:tHj}

As a proper test of SPINUP we include an unfolding application to parton level. An
established, challenging example process is associated single-top and
Higgs production with a purely hadronic
signal~\cite{Butter:2022vkj, Heimel:2023mvw},
\begin{align}
  p p \to t H j \to (b jj) \; (\gamma \gamma) \; j + \text{jets}\; ,
  \label{eq:proc}
\end{align}
We allow for a complex top Yukawa
coupling~\cite{Artoisenet:2013puc} with a CP-phase $\alpha$,
\begin{align}
  \lag_{t\bar{t}H} = - \frac{y_t}{\sqrt{2}}
  \Big[ a\cos \alpha \; \left( \bar{t}t \right) + 
  ib\sin \alpha \; \left( \bar{t}\gamma_5 t \right) \Big] \; H \; ,
\label{eq:LttH}
\end{align}
with $a = 1$ and $b=2/3$~\cite{Demartin:2015uha}, keeping the cross
section for $gg\to H$ constant.

We define the parton level as the top, Higgs and forward jet before
decays and QCD radiation, defining seven independent phase space
dimensions. From there we apply the usual simulation chain with
decays, parton shower and detector simulation, as described in
Ref.~\cite{Butter:2022vkj}. At reco-level, we consider the momenta of
the two photons, one b-tagged jet and the three light jets with the
largest transverse momentum. We make a range of kinematic cuts, following Ref.~\cite{Butter:2022vkj}.

\begin{figure}[!b]
    \includegraphics[width=0.325\textwidth,page=5]{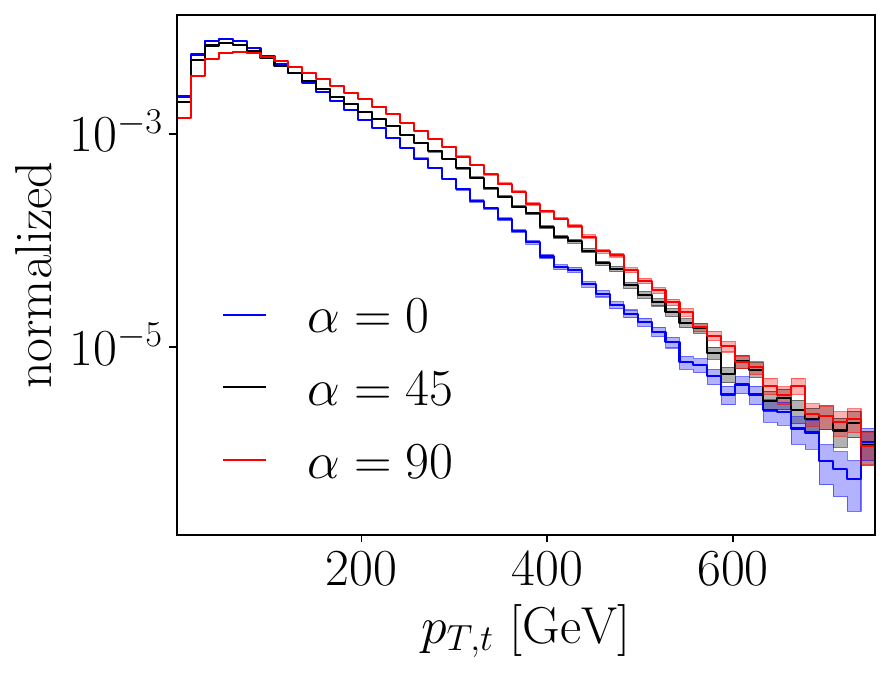}
    \includegraphics[width=0.325\textwidth,page=7]{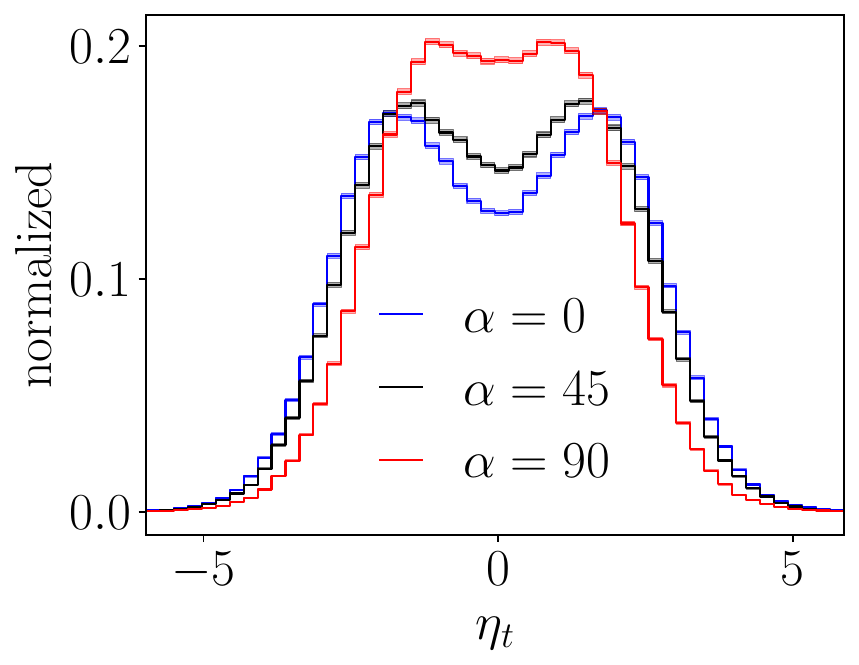}
    \includegraphics[width=0.325\textwidth,page=15]{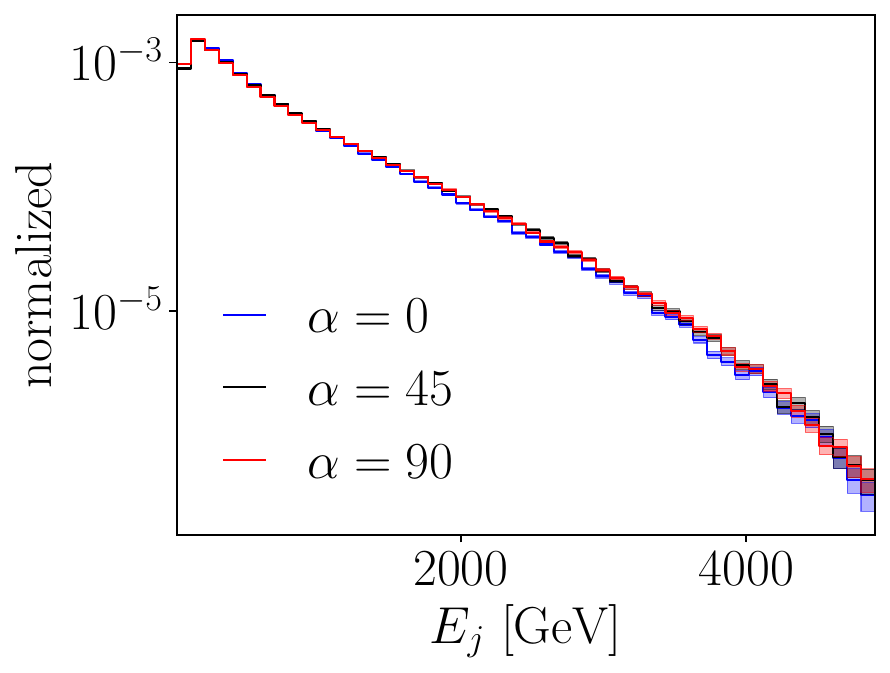} \\
    \includegraphics[width=0.325\textwidth,page=24]{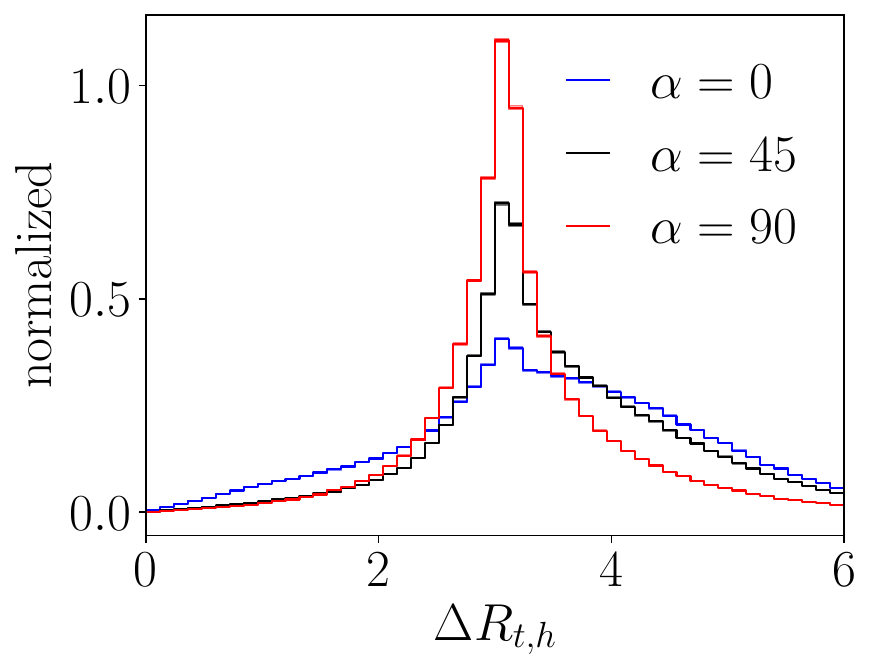}
    \includegraphics[width=0.325\textwidth,page=26]{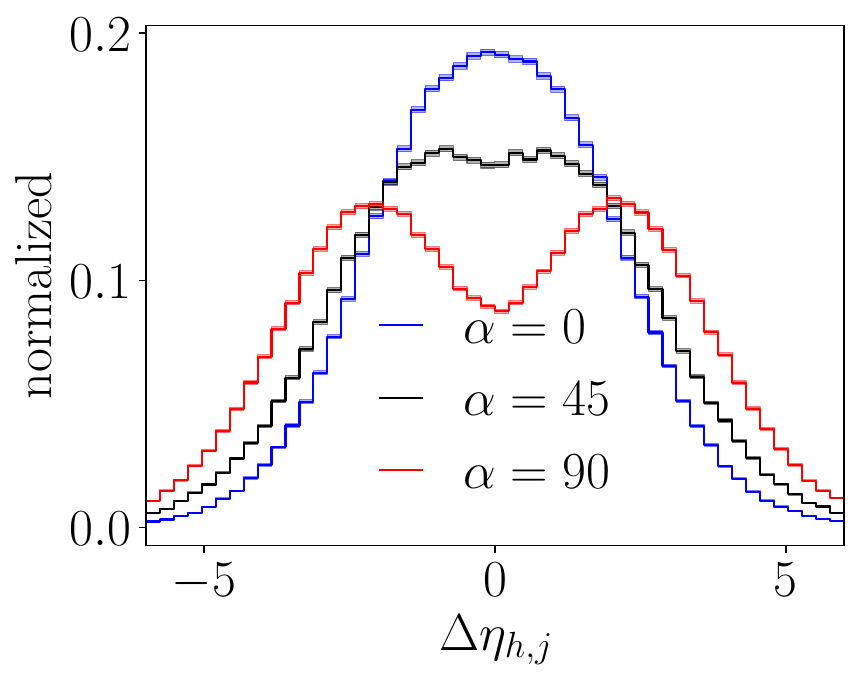}
    \includegraphics[width=0.325\textwidth,page=22]{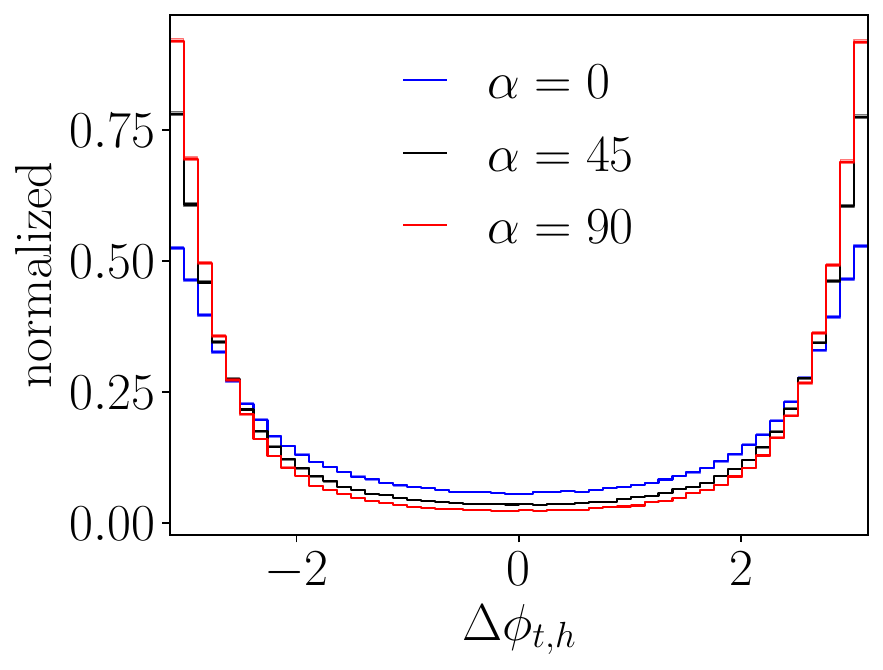}
    \caption{Dependence of the parton-level distribution on the parameter $\alpha$}
    \label{fig:MEM_parton}
\end{figure}

The model dependence of the simulation is defined by the choice of $\alpha$. The
parton-level momenta contain the complete information of the forward
process, resulting in a model-independent transfer function. Looking
at the parton-level distributions for $\alpha=0^\circ,45^\circ,90^\circ$ in
Fig.~\ref{fig:MEM_parton}, we observe that the angular separations are
particularly sensitive. To ensure a good coverage of phase space for
different CP-angles during the training of our transfer function, we
generate 4.5M events with $\alpha$ sampled uniformly between
$-180^\circ$ and $180^\circ$. For the training of the unfolding
network we use 450K `measured' reco-level events with
$\alpha=45^\circ$.  The goal is to unfold to the parton level such
that $\alpha$ can be measured there without any training bias.

\paragraph{Transfer function}
We first look at the trained transfer function to confirm that
$p(\xr|\xp)$ generalizes between different $\alpha$ values.  As
mentioned above, the training data is sampled uniformly between
$-180^\circ$ and $+180^\circ$, without conditioning. The
hyperparameters can be found in Tab.~\ref{tab:mem_hyperparameters} in the appendix. In
Fig.~\ref{fig:MEM_transfer} we show a set of reco-level distributions
when applying the transfer network to parton-level distributions with
$\alpha=0^\circ, 45^\circ, 90^\circ$. In all cases the correct
reco-level distributions are recovered, indicating no prior dependence
in the forward mapping. In the $\Delta R_{j_b, j_1}$ distribution the
network interpolates through the jet algorithm cut at $\Delta R_{j_b,
  j_1}=0.4$. Such a shortcoming in the learned transfer probability
will inherently limit the precision of the unfolding. However, in our 
case it only
affects a sub-leading feature, and we do not observe any general
problems.

\begin{figure}[!t]
    \includegraphics[width=0.49\textwidth,page=7]{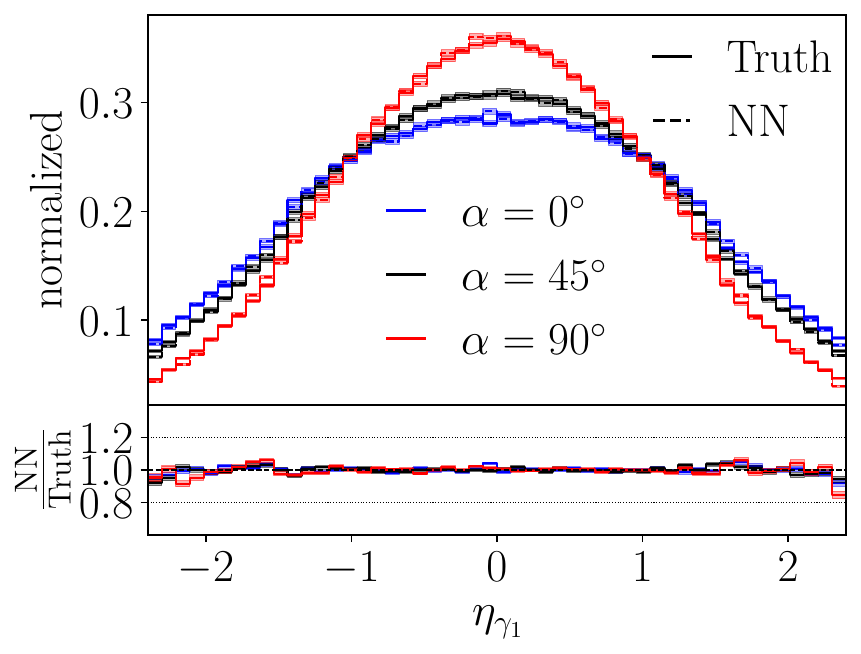}
    \includegraphics[width=0.49\textwidth,page=23]{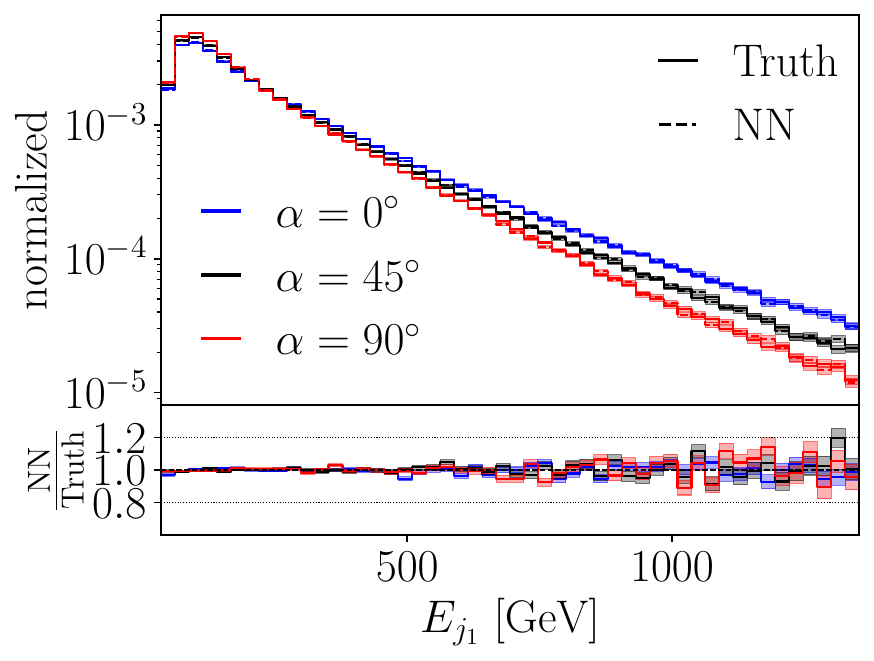} \\
    \includegraphics[width=0.49\textwidth,page=26]{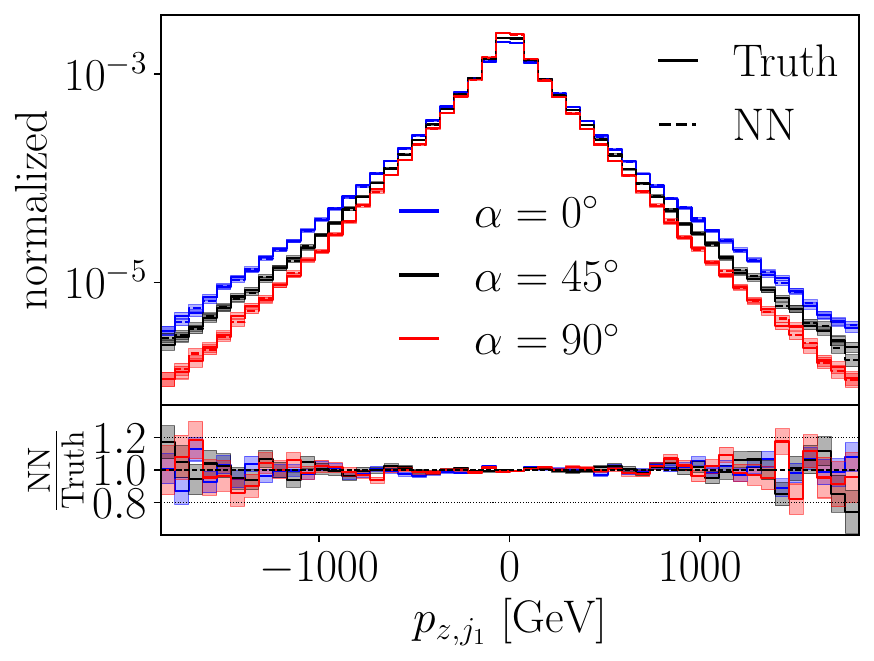} 
    \includegraphics[width=0.49\textwidth,page=55]{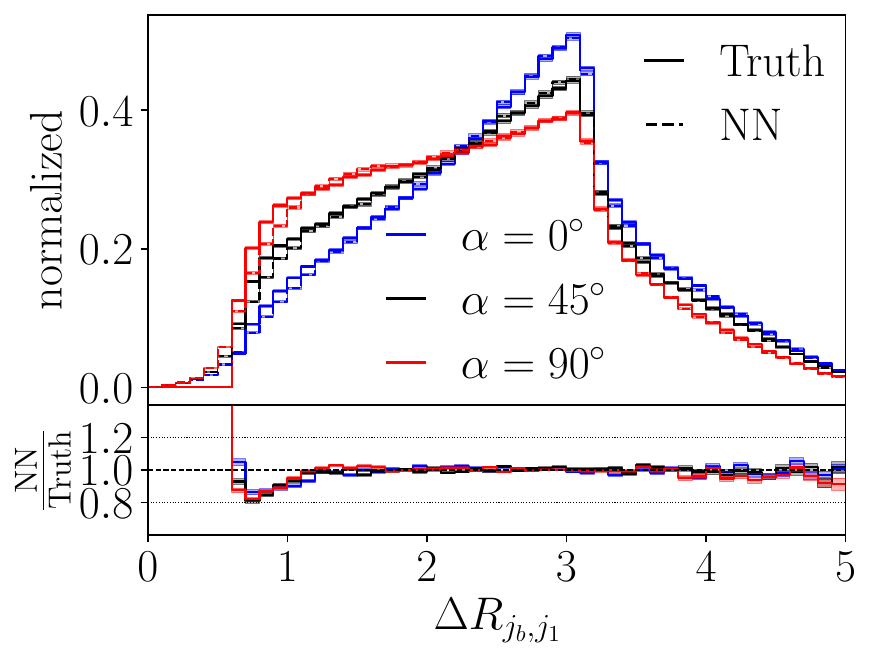} 
    \caption{Transfer network trained on the $tHj$ parton-level dataset}
    \label{fig:MEM_transfer}
\end{figure}

\subsection{Unfolding}

\begin{figure}[!t]
    \includegraphics[width=0.49\textwidth,page=5]{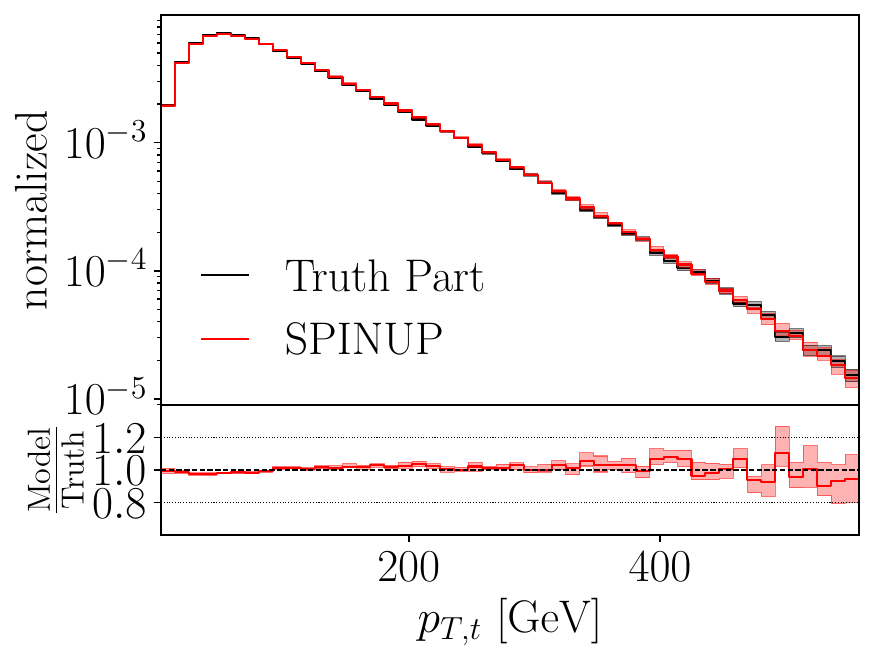}
    \includegraphics[width=0.49\textwidth,page=7]{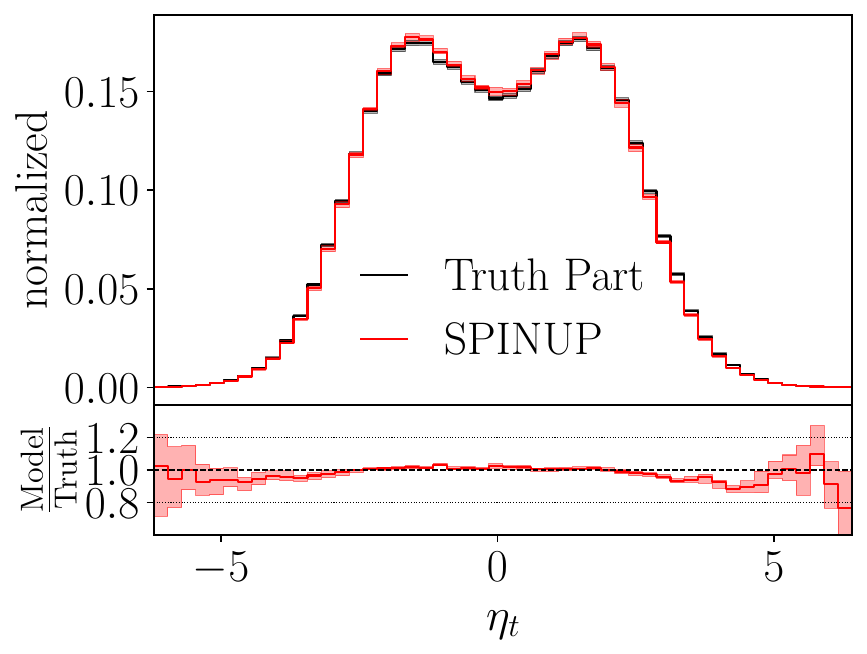} \\
    \includegraphics[width=0.49\textwidth,page=15]{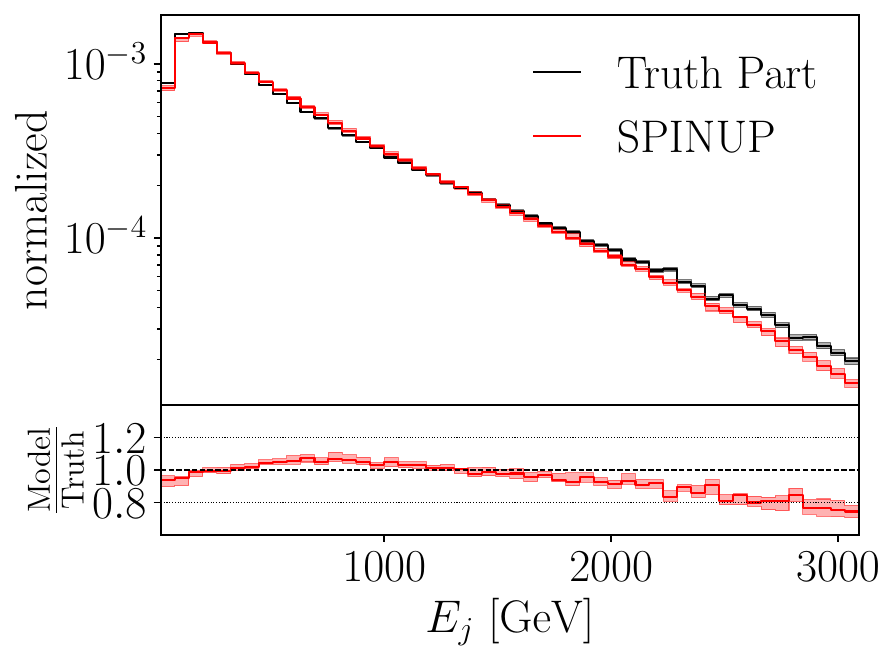} 
    \includegraphics[width=0.49\textwidth,page=24]{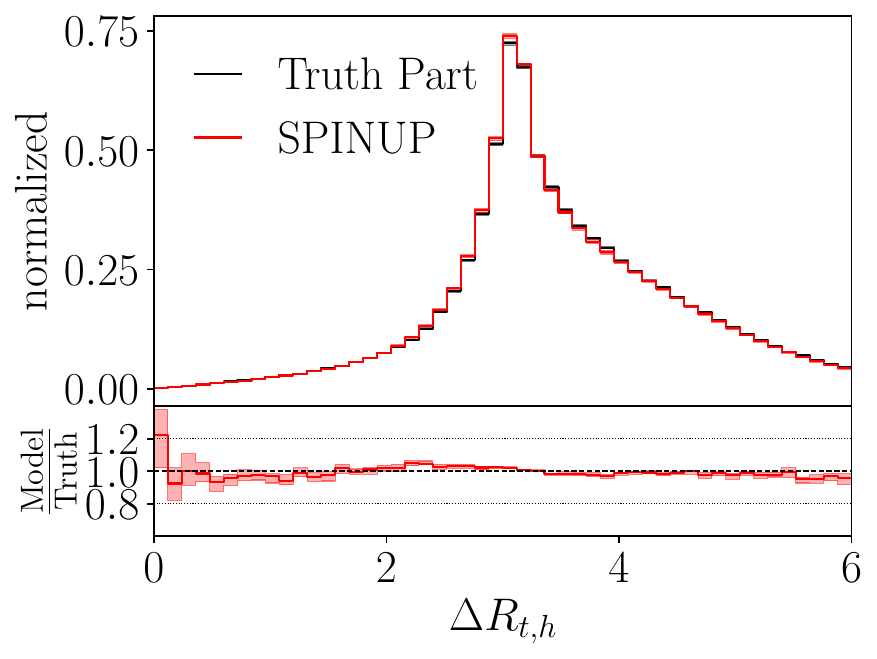} \\
    \includegraphics[width=0.49\textwidth,page=26]{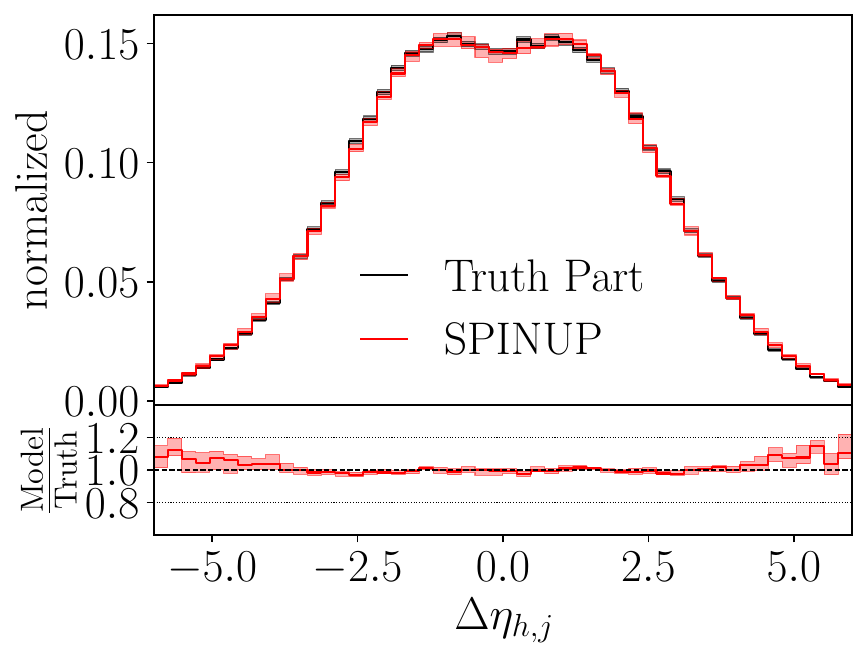}
    \includegraphics[width=0.49\textwidth,page=22]{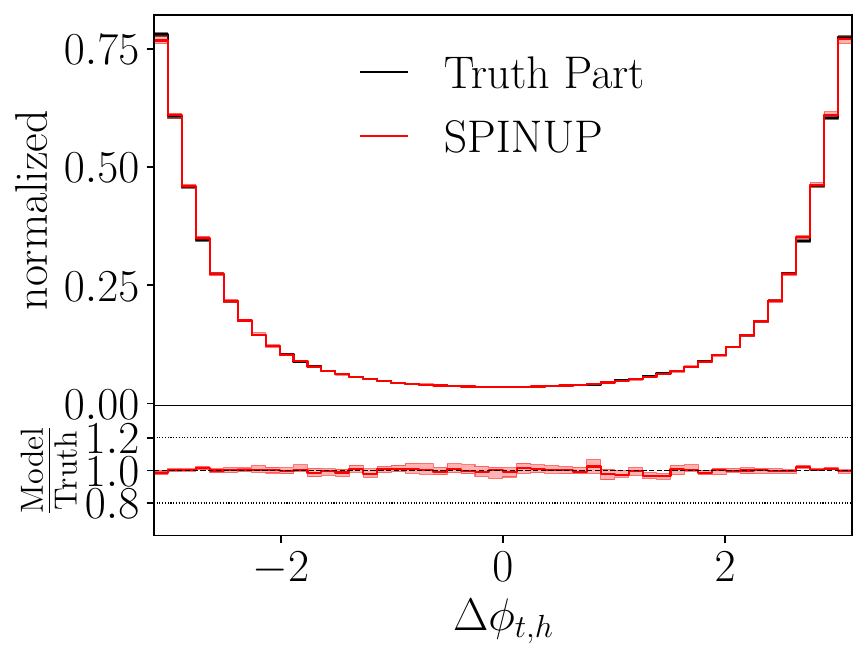}
    \caption{Unfolded observables at parton-level using SPINUP for $\alpha=45^\circ$.}
    \label{fig:MEM_parton_45}
\end{figure}

We illustrate the SPINUP performance for $\alpha=45^\circ$. Again, we
train an ensemble of networks and show the median and 68\%
quantiles. The training hyperparameters can be found in the appendix
in Tab.~\ref{tab:mem_hyperparameters}. Some unfolded distributions at parton level are
shown in Fig.~\ref{fig:MEM_parton_45}. Deviations between truth and the unfolded distribution are at the percent-level over most of the phase space. It shows no sign
of prior dependence and recovers the parton-level distribution for
$\alpha=45^\circ$. The observed reco-level distributions contain
enough information to reconstruct the correct parton-level
distribution and the method delivers this task.

The variation over the ensemble is very small over most of phase space
and only mildly grow towards the tails, indicating that the method
does not find significant degeneracy in the solution space. Notably,
in $E_j$ the tail towards large energies is consistently
underestimated by all ensemble members.

\begin{figure}[!t]
    \includegraphics[width=0.49\textwidth,page=1]{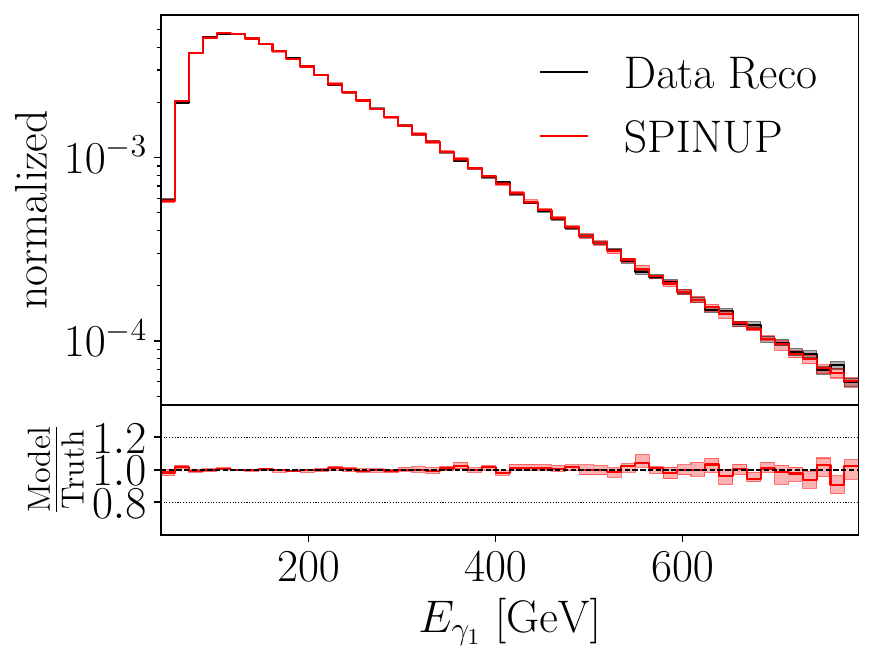}
    \includegraphics[width=0.49\textwidth,page=27]{figs/mem/mem_observables_reco.pdf}\\
    \includegraphics[width=0.49\textwidth,page=7]{figs/mem/mem_observables_reco.pdf}
    \includegraphics[width=0.49\textwidth,page=23]{figs/mem/mem_observables_reco.pdf}\\
    \includegraphics[width=0.49\textwidth,page=27]{figs/mem/mem_observables_reco.pdf}
    \includegraphics[width=0.49\textwidth,page=55]{figs/mem/mem_observables_reco.pdf} \\
    \caption{Reco-level observables for $\alpha=45^\circ$, obtained by passing the SPINUP unfolding results through the forward model.}
    \label{fig:MEM_reco_45}
\end{figure}

We move on to the results at reco-level, obtained by passing the
learned SPINUP unfolding through the forward network. Results for
$\alpha=45^\circ$ are shown in Fig.~\ref{fig:MEM_reco_45}.  Reproducing the
observed reco-level data is the training objective of the unfolding
network. SPINUP faithfully reproduces the reco-kinematics with high precision, keeping differences between truth and unfolded distribution at the 
percent-level over the entire phase space. Again, our method
does not show any signs of prior dependence. We observe reco-level data
for a specific choice of $\alpha$ and the method is trained to
reproduce this observed distribution. In the $\Delta R$ observables
the mis-modelling at $\Delta R=0.4$ remains, because like any other
density estimation neural network, SPINUP struggles to model hard
cuts.

\subsection{Inference}

After demonstrating that SPINUP unfolds to partons at the per-cent
level and without prior dependence, we test whether we can measure $\alpha$ from unfolded, public data.  At parton level we have direct
analytic access to the likelihood via the differential cross section
\begin{align}
    p(\xp|\alpha) = \frac{1}{\sigma (\alpha)} \frac{d\sigma(\alpha)}{d\xp} \; .
    \label{eq:diff_cross_section}
\end{align}
This means that we can evaluate the theory likelihood of an event
under a parameter hypothesis $\alpha$. For a sample of parton level
events $\{x_\text{part, i}\}$, we can calculate~\cite{Cranmer:2006zs}
\begin{align}
    \loss(\alpha) = \sum_i^{N_\text{data}} \log p(x_\text{part, i}|\alpha) 
\end{align}
After unfolding, this inference step is technically
trivial. We evaluate the likelihood on a
fine grid of $\alpha$ values and extract the maximum.

\paragraph{Acceptance effects}
One subtlety that we have to account for when using
the differential cross section as in Eq.\eqref{eq:diff_cross_section} is event acceptance. We apply an event selection at reco-level, which in turn leads to a smeared selection at parton-level. This is encoded in the efficiency $\epsilon (\xp)$, defined as the probability to accept a given part-level event after forward simulation.
All components of SPINUP are trained on the fiducial phase space after cuts, corresponding to the total cross section
\begin{align}
    \sigma_\text{fid} (\alpha) = \sigma (\alpha) \int \d \xp \; \epsilon (\xp) \; p(\xp | \alpha) \; .
    \label{eq:cross_section_corrected}
\end{align}
This changes the learned or unfolded density over parton-level phase space to
\begin{align}
    p(\xp|\alpha) = \frac{1}{\sigma_\text{fid} (\alpha)} \frac{d\sigma(\alpha)}{d\xp} \epsilon (\xp)  \; ,
    \label{eq:diff_cross_section_corrected}
\end{align}
This effect is discussed in detail in Ref.~\cite{Heimel:2023mvw}. Following the same strategy we train a binary classifier $\epsilon_\theta (\xp)$ which predicts reco-level acceptance as a function of the parton-level kinematics. It will directly give us the acceptance probability and allow us to correct the densities. 

\paragraph{Results}
To measure the CP-phase $\alpha$ from unfolded data, we generate samples from the learned unfolding network $\xp \sim p_\text{unfold}(\xp)$ and calculate the parton-level log-likelihood on a grid of 1000 values $\alpha_\text{true}\pm15$. We then extract the maximum-likelihood solution as the minimum of the resulting likelihood parabolas. We estimate the uncertainty by repeating this procedure 20 times. 

The resulting error on $\alpha$ scales with the number of sampled events. Since we have encoded the unfolded distribution in an unconditional generative network $p_\theta(\xp)$, we can generate an arbitrary number of samples. Doing this with a single neural network sends the estimator uncertainty to zero. This is because this uncertainty only represents the MC approximation uncertainty of sampling a generative model rather than evaluating an explicit density. Instead, we have to take into account the uncertainty on the unfolded distribution itself, which is addressed by incorporating the spread from the trained ensemble. Instead of resampling the same network 20 times, we draw one sample of events from each of the 20 ensemble members. This gives us 20 estimates for $\alpha$. 

\begin{figure}[t]
    \includegraphics[width=0.495\textwidth]{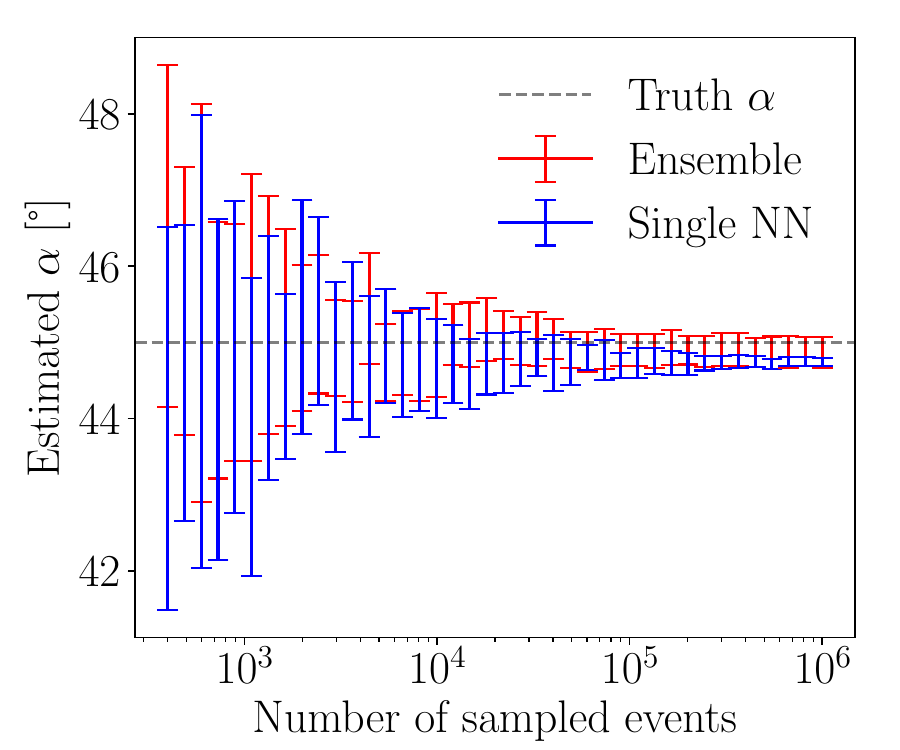}
    \includegraphics[width=0.495\textwidth]{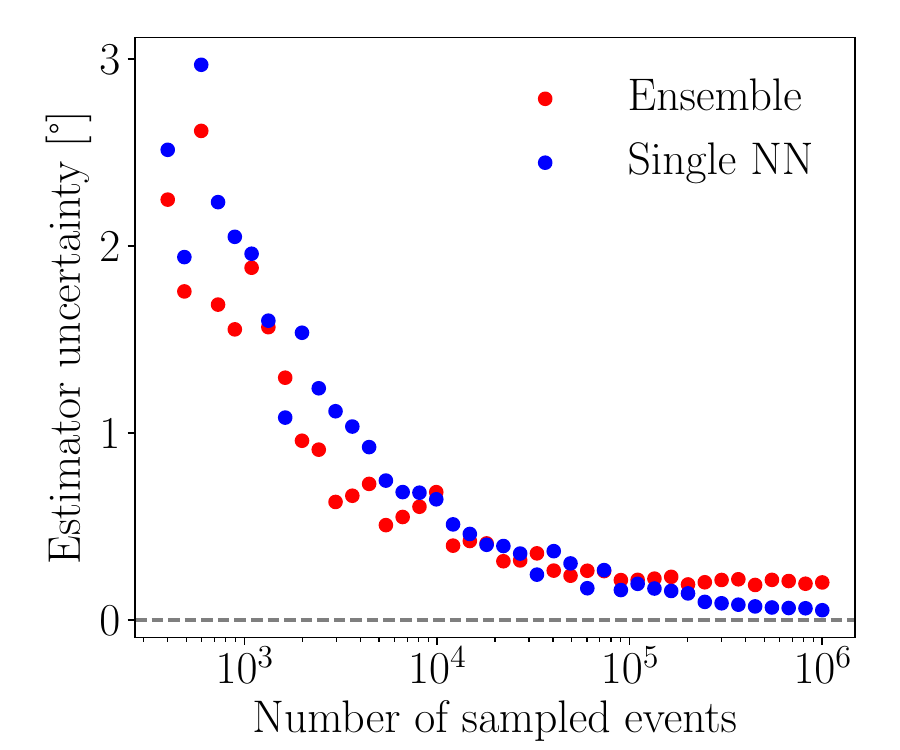} \\
    \caption{Parton-level parameter inference on the unfolded distribution for the $\alpha=45^\circ$ event sample. The uncertainties are the standard deviations over 20 event samples, either from the 20 different networks in the ensemble or from the same network repeatedly.}
    \label{fig:inference_mem}
\end{figure}

The results from this procedure are shown in Fig.~\ref{fig:inference_mem}. The left panel shows the estimated $\alpha$ using either 20 generated events sampled from a single network, or one event sample from each of the 20 networks in the full ensemble. The right panel shows the scaling of the uncertainty with the number of sampled events. Indeed, the unfolded distribution gives a correct measurement of $\alpha$, confirming  that our SPINUP method has recovered the truth distribution.

\begin{figure}[!b]
    \includegraphics[width=0.495\textwidth]{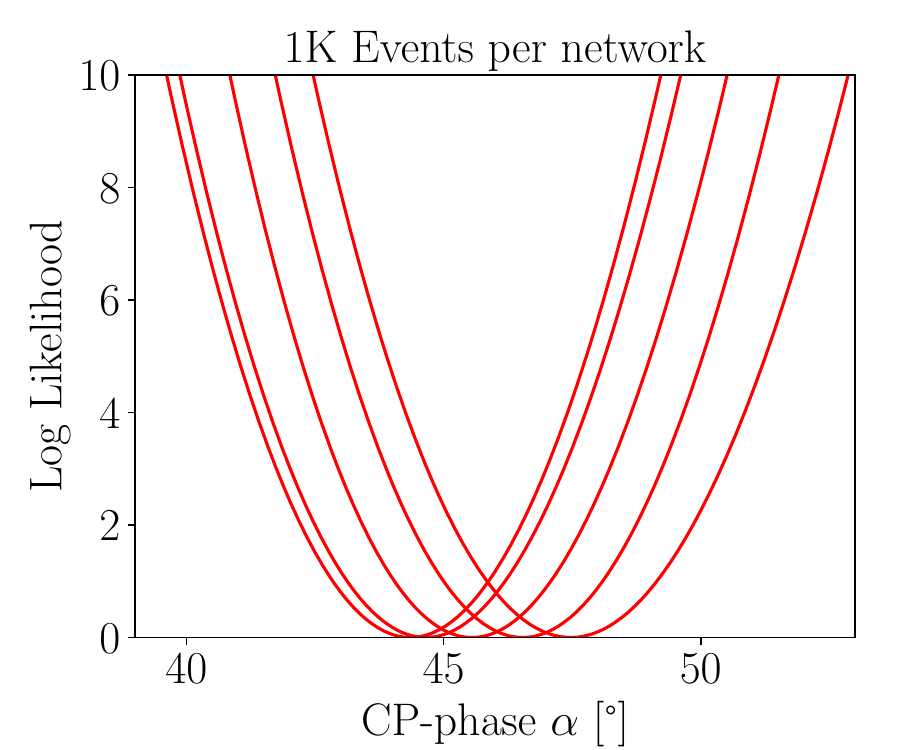}
    \includegraphics[width=0.495\textwidth]{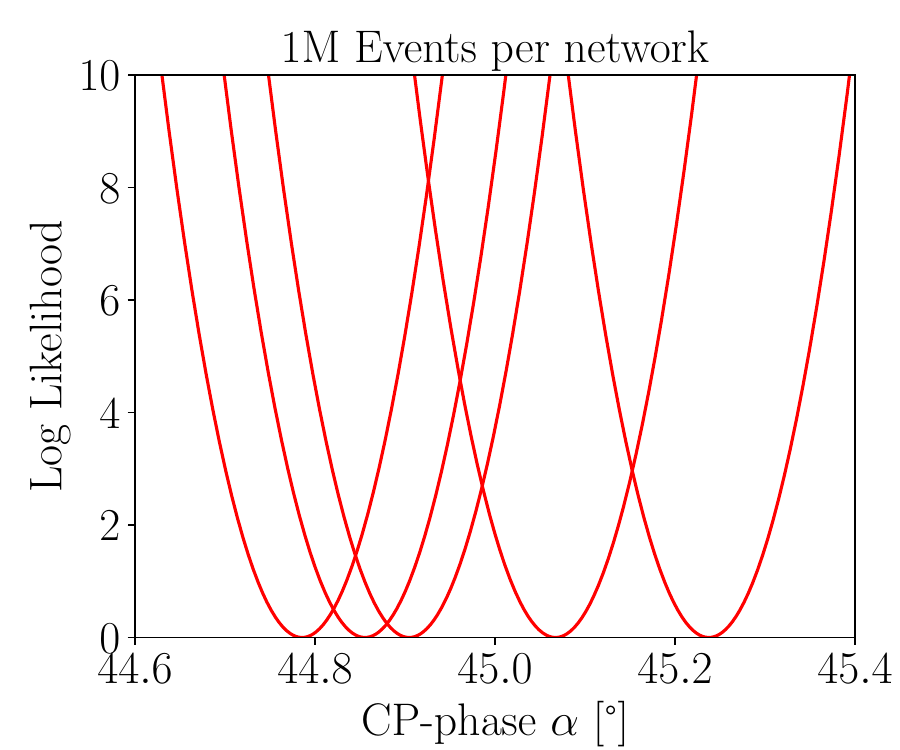} \\
    \caption{Likelihood parabolas from 5 networks from the SPINUP ensemble. The left plot shows results from 1K events sampled per network, the right plot shows results for sampling 1M events per network.}
    \label{fig:parabolas_mem}
\end{figure}

Even using only 400 events from the sampled unfolded distribution, $\alpha$ can already be estimated, albeit with a large uncertainty. In the low event number regime, the single network measurement and the ensemble measurement have similar uncertainties, indicating that they are dominated by the statistical fluctuations. When we increase the number of sampled events, the uncertainties decrease and the difference between the ensemble and the single network becomes visible. For the single network the uncertainties vanish, and no longer cover the true value for more than $10^5$ samples. For the ensemble measurement the uncertainties reach a plateau when the statistical uncertainties on the individual estimates become negligible and we are  dominated by the ensemble uncertainty. The ensemble translates the statistical uncertainty from a finite number of data events into a systematic uncertainty on the measured parameter, which covers the true value.

Alternatively, we can look at the log-likelihood parabolas extracted from 5 networks from the SPINUP ensemble shown in Fig.~\ref{fig:parabolas_mem}. The left panel shows parabolas calculated with 1K events sampled from the networks. They are scattered around the true $\alpha=45^\circ$, but due to the small number of sampled events they are very broad and do not allow a precise measurement. The uncertainty is dominated by the statistical uncertainties of the individual estimates. For large statistics with 1M sampled events, shown in the right panel, the individual parabolas become very narrow and the uncertainty is dominated by the spread of the minima positions. 

\subsection{Lower data statistic}
Finally, we investigate the behavior of SPINUP when training the algorithm on a lower number of measured reco-level events. We leave the number of simulated events used to train the transfer and NIS networks invariant. This allows us to emulate the realistic scenario of unfolding a process with a small fiducial cross section, such as $tHj$.

\begin{figure}[!t]
    \includegraphics[width=0.495\textwidth]{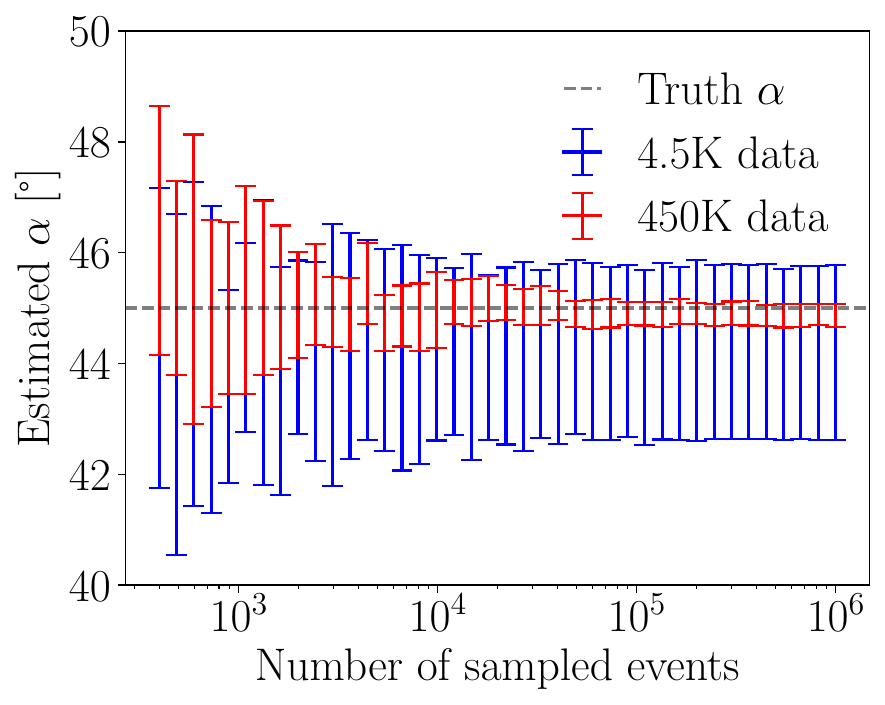}
    \includegraphics[width=0.495\textwidth]{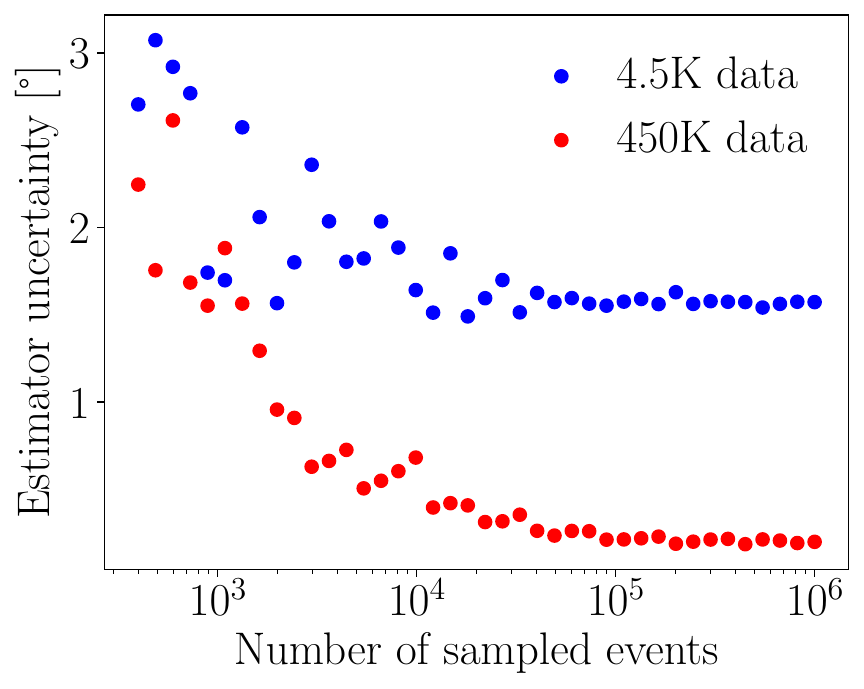} \\
    \caption{Parton-level parameter inference on the unfolded distribution for the $\alpha=45^\circ$ event sample. The uncertainties are derived from the network ensemble.}
    \label{fig:low_statistics_mem}
\end{figure}

In Fig.~\ref{fig:low_statistics_mem} we compare the parameter estimation results from training SPINUP on either the full 'measured'  dataset (450K events) or just 1 percent of that dataset (4.5K events). We observe that even for just 4.5K measured events, SPINUP allows to accurately recover the theory parameter from the unfolded distribution. The uncertainty in the parameter estimation is significantly larger, as we would expect.

If the ensembling uncertainty captured purely the statistical uncertainty from limited training data, we would expect it to scale as $\frac{1}{\sqrt{n}}$. However, the ensemble also captures other sources of uncertainty, such as stochastic network initialization and training, and potential degeneracy in the solution space. In this preliminary experiment we observe the factor to be almost $\frac{1}{\sqrt{n}}$, indicating that the limited statistics are the leading source of uncertainty. We leave a more systematic study of the uncertainties for future work.

\section{Outlook}
\label{sec:Outlook}

Modern machine learning makes it possible to
precisely unfold high-dimensional, unbinned
spaces, relying on reweighting,
distribution mapping, or posterior estimation. SPINUP 
instead estimates the forward
mapping and 
extracts the unfolded
distribution by proposing part-level configurations with neural
importance sampling. The forward information loss is 
captured with ensembling. Technically, SPINUP combines 
a forward transfer network, a NIS network, and the generative
unfolding network.

We first illustrated SPINUP for Gaussian toy models, where
it  
extracts the part-level truth, clearly separated from the
part-level simulation used to train the transfer
network. The reco-level data can be recovered with a 
convolution with the transfer network. 
This remains true even for a larger smearing where the
reco-level for simulation and truth become very similar, and the
ensembling captures the increased uncertainty of the unfolded
part-level distribution. In the example of a bimodal Gaussian
distribution with a peak at positive and negative values, and a
forward process that loses the sign information, we find that the
ensemble maps out the space of possible part-level distributions that
map to the same reco-level distributions, again confirming the
effectiveness of our ensembling procedure.

Next, we apply our method to the OmniFold
dataset~\cite{Andreassen:2019cjw}, a widely used benchmark dataset for
machine-learning based unfolding methods containing a set of six
jet-substructure observables in Z+jets production. Using events
generated by Pythia as simulation and events generated by Herwig as
truth, we find that the unfolded distribution does not recover the
truth, with deviations up to $10\%$. However, this is not due to a
lack of precision of our unfolding method. Instead, the
model-independence of the forward mapping is lost in the reduction of
the high-dimensional space of jet-constituent four-momenta to
jet-substructure observables, violating a fundamental assumption of
unfolding. We demonstrate that our method is able to recover the truth
distribution at the percent level for a synthetic reco-level dataset
sampled using our learned transfer function.

Finally, we use SPINUP to unfold associated single-top and Higgs
production events to parton level. We allow for mixed CP-even and
CP-odd interactions in the top Yukawa coupling with a CP-phase
$\alpha$. We train our transfer network on events with a uniformly
distributed CP-phase and then unfold reco-level events with
$\alpha=45^\circ$, recovering most parton-level distributions with
percent-level precision. We then use the analytic expression for the
matrix element to extract a log-likelihood for $\alpha$ from the
unfolded data and correctly infer the parameter. We estimate the
uncertainty arising from our unfolding procedure by repeating the
inference for an ensemble of unfolding networks. Moreover, we show
that the SPINUP method remains robust for very low numbers of measured
events.

Altogether, the SPINUP method is a valuable addition to the growing
selection of ML-based unfolding methods as it is simulation-prior
independent by construction and is suitable
even for applications with large losses of information in the forward
process and low numbers of unfolded events.

\section*{Acknowledgments}
We thank Sofia Palacios Schweitzer for very fruitful discussions, and for simulating the top decay dataset. NH is supported by the BMBF Junior Group
Generative Precision Networks for Particle Physics (DLR 01IS22079). MK is supported by the US Department of Energy (DOE) under grant
DE-AC02-76SF00515. TH is supported by the PDR-Weave grant FNRS-DFG numéro T019324F (40020485).
This work was supported by the DFG under grant 396021762 -- TRR~257: \textsl{Particle Physics Phenomenology after the   Higgs Discovery}, and through Germany's Excellence Strategy EXC~2181/1 -- 390900948 (the \textsl{Heidelberg STRUCTURES Excellence Cluster}). The authors acknowledge support by the state of Baden-Württemberg through bwHPC
and the German Research Foundation (DFG) through grant INST 35/1597-1 FUGG.

\clearpage
\appendix

\section{Hyperparameters}
\begin{table}[ht!]
    \centering
    \begin{small} \begin{tabular}[t]{l|ccc}
    \toprule
    Parameter  & Unfolding-NN \qquad& Transfer-NN \qquad & Sampling-NN \\
    \midrule
    Network&  Transfermer & Transfermer & INN  \\
    Optimizer& Adam & Adam & Adam \\
    Learning rate & 0.001 & 0.0003 & 0.0003 \\
    LR schedule & Cosine annealing & One cycle & One cycle\\
    Batch size &  512 & 1024 & 1024\\
    Epochs & 60 & 200 & 200 \\
    INN Blocks & - & - & 6 \\
    Subnet layers & 5 & 5 & 5 \\
    Subnet dim & 256 & 256 & 256 \\
    RQS bins & 20 & 20 & 20 \\
    Transformer blocks & 4 & 4 & - \\
    Transformer heads & 4 & 4 & - \\
    Embedding dim & 64 & 64 & - \\
    Feedforward dim & 256 & 256 & - \\
    MC samples & 64 & - & - \\
    Pretrain epochs & 200 & - & - \\
    \bottomrule
    \end{tabular} \end{small}
    \caption{Network and training hyperparameters for the jet substructure dataset in Sec.~\ref{sec:omnifold}}
    \label{tab:6d_hyperparameters}
\end{table}

\begin{table}[ht!]
    \centering
    \begin{small} \begin{tabular}[t]{l|ccc}
    \toprule
    Parameter  & Unfolding-NN \qquad& Transfer-NN \qquad & Sampling-NN \\
    \midrule
    Network&  Transfermer & Transfermer & INN  \\
    Optimizer& Adam & Adam & Adam \\
    Learning rate & 0.001 & 0.0003 & 0.0003 \\
    LR schedule & Cosine annealing & One cycle & One cycle\\
    Batch size &  512 & 1024 & 1024\\
    Epochs & 30 & 350 & 350 \\
    INN Blocks & - & - & 20 \\
    Subnet layers & 5 & 5 & 5 \\
    Subnet dim & 256 & 256 & 256 \\
    RQS bins & 30 & 30 & 30 \\
    Transformer blocks & 6 & 6 & - \\
    Transformer heads & 6 & 6 & - \\
    Embedding dim & 64 & 64 & - \\
    Feedforward dim & 256 & 256 & - \\
    MC samples & 64 & - & - \\
    Pretrain epochs & 50 & - & - \\
    \bottomrule
    \end{tabular} \end{small}
    \caption{Network and training hyperparameters for the parton level $tHj$ dataset in Sec.~\ref{sec:tHj}}
    \label{tab:mem_hyperparameters}
\end{table}

\clearpage
\section{Additional results}
\subsection*{Jet substructure dataset}
\subsubsection*{Transfer model}
\begin{figure}[!h]
    \includegraphics[width=0.32\textwidth, page=1]{figs/omnifold/observables_omnifold_transferfunction.pdf}
    \includegraphics[width=0.32\textwidth, page=2]{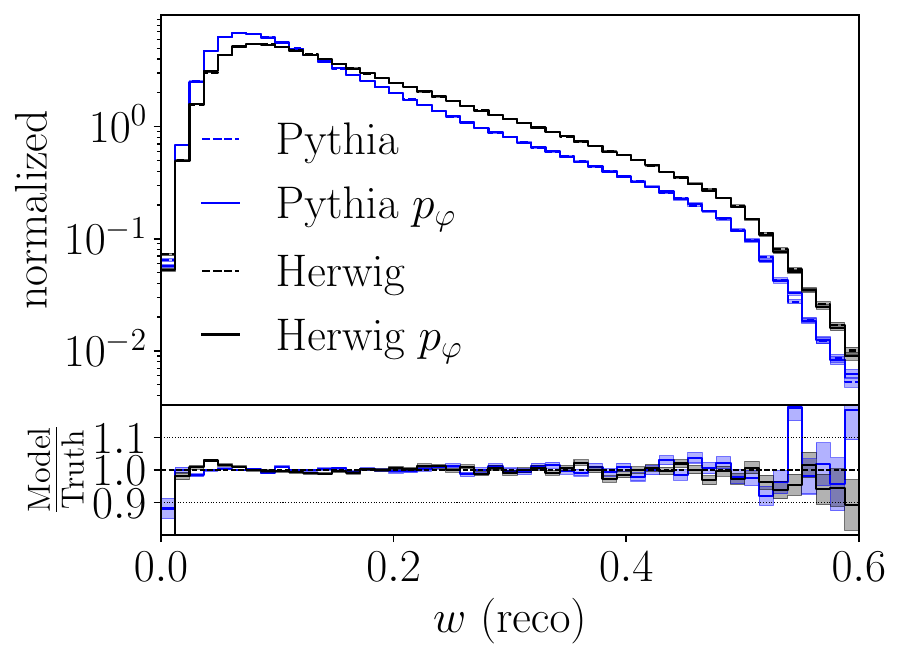}
    \includegraphics[width=0.32\textwidth, page=3]{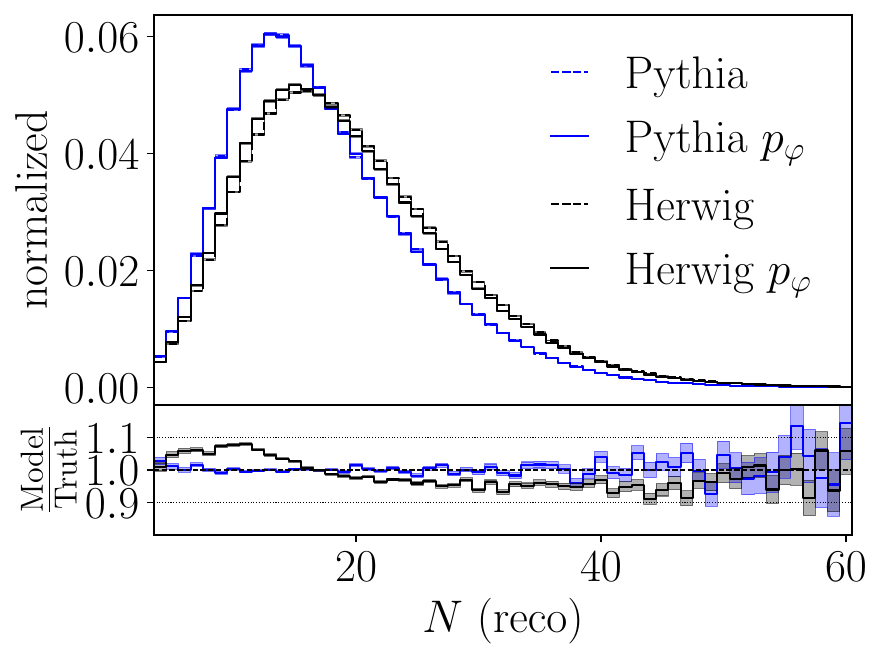}\\
    \includegraphics[width=0.32\textwidth, page=4]{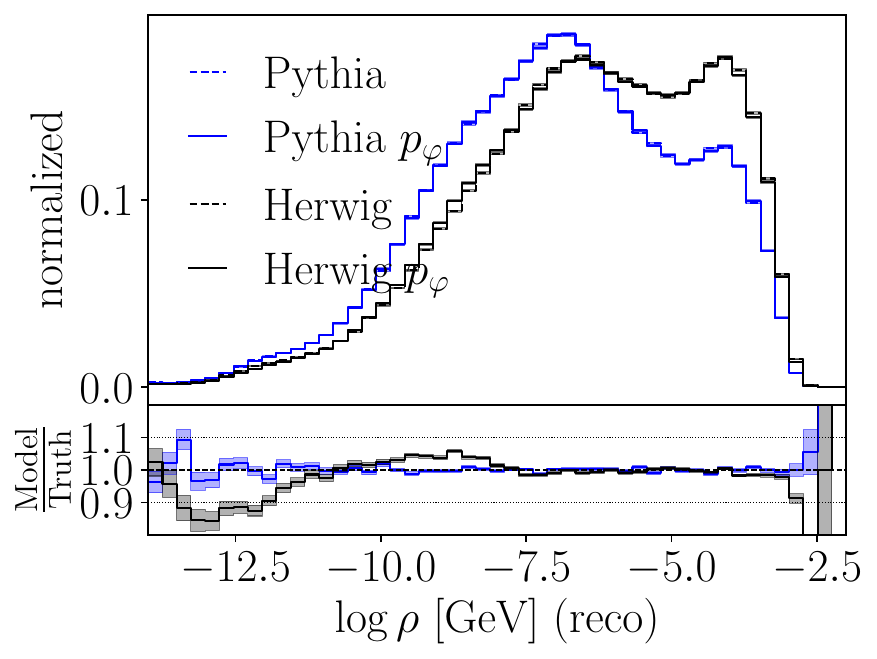}
    \includegraphics[width=0.32\textwidth, page=5]{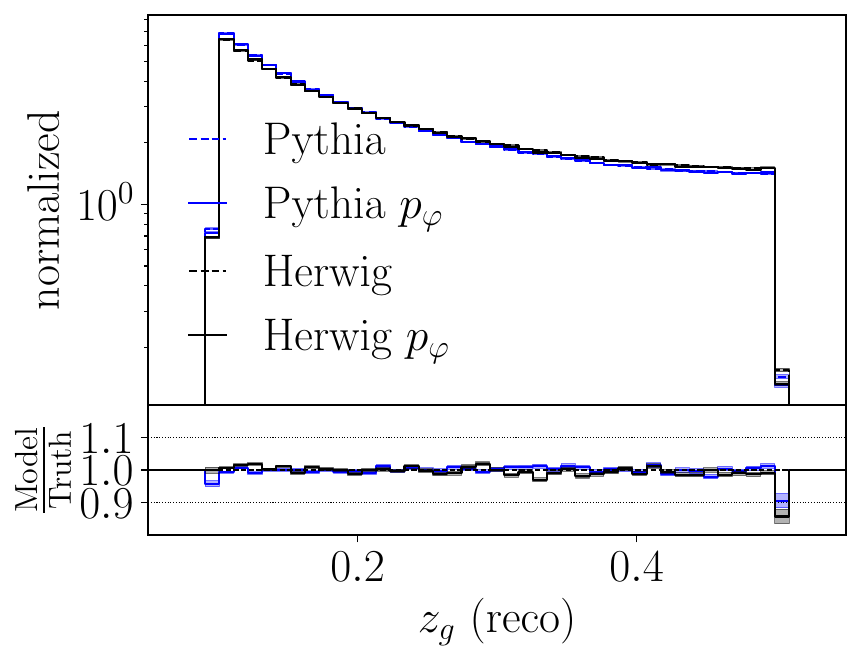}
    \includegraphics[width=0.32\textwidth, page=6]{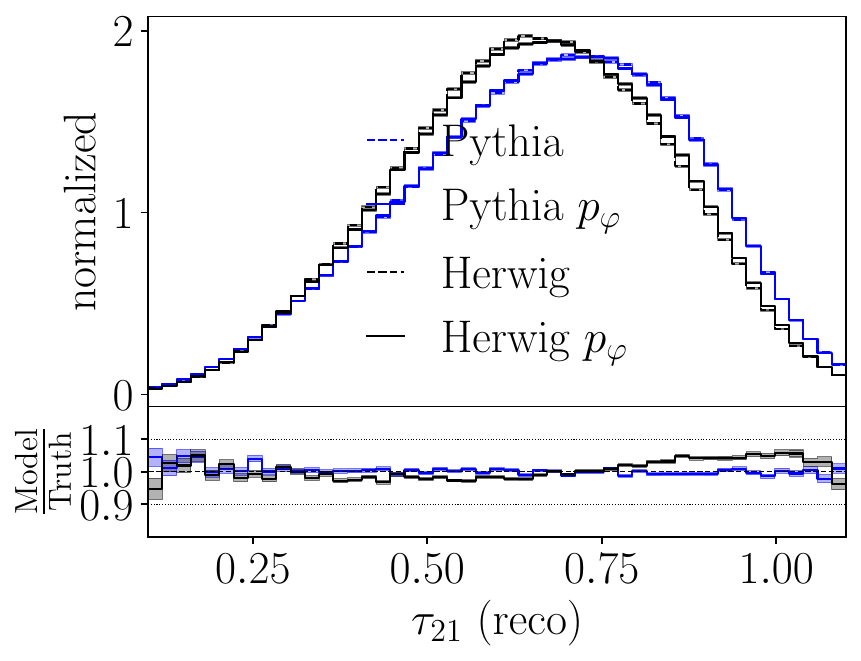}
    \caption{Full set of plots for the transfer model trained on Pythia and applied to Herwig.}
    \label{fig:omnifold_transfer_full}
\end{figure}

\subsubsection*{Unfolding results on original Herwig}
\begin{figure}[!h]
    \includegraphics[width=0.32\textwidth, page=1]{figs/omnifold/observables_part_omnifold_trueHerwig.pdf}
    \includegraphics[width=0.32\textwidth, page=2]{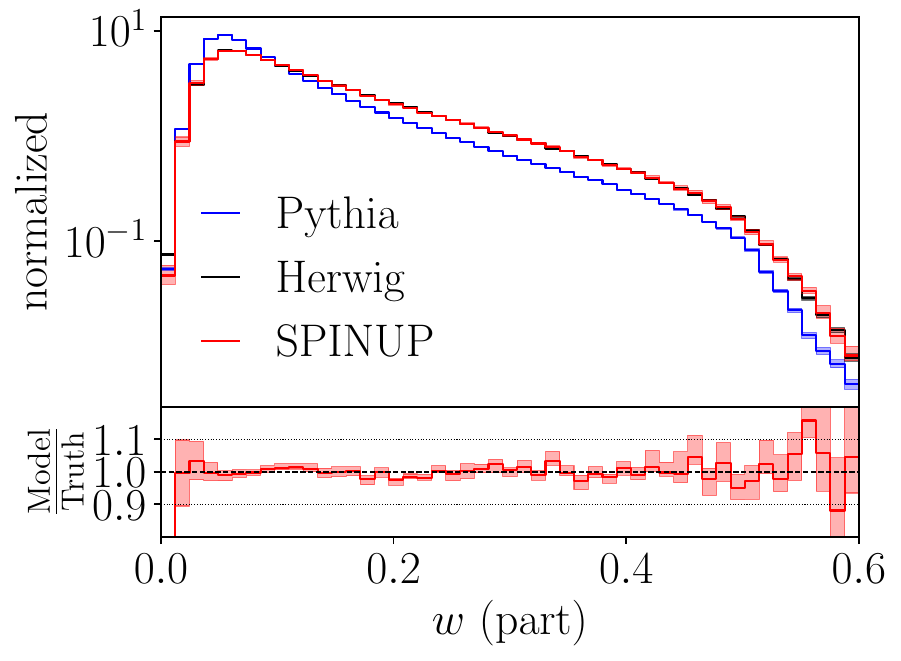}
    \includegraphics[width=0.32\textwidth, page=3]{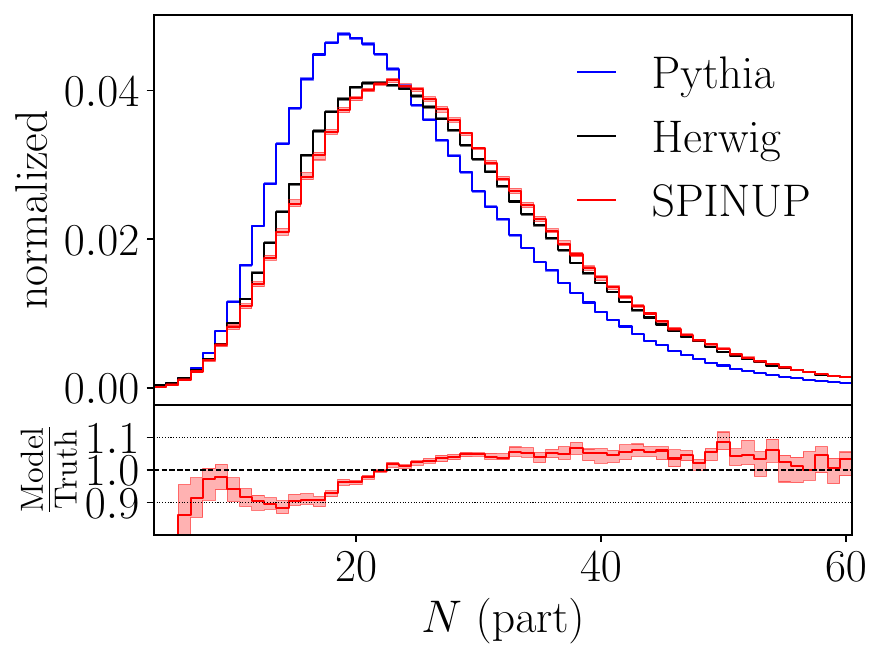}\\
    \includegraphics[width=0.32\textwidth, page=4]{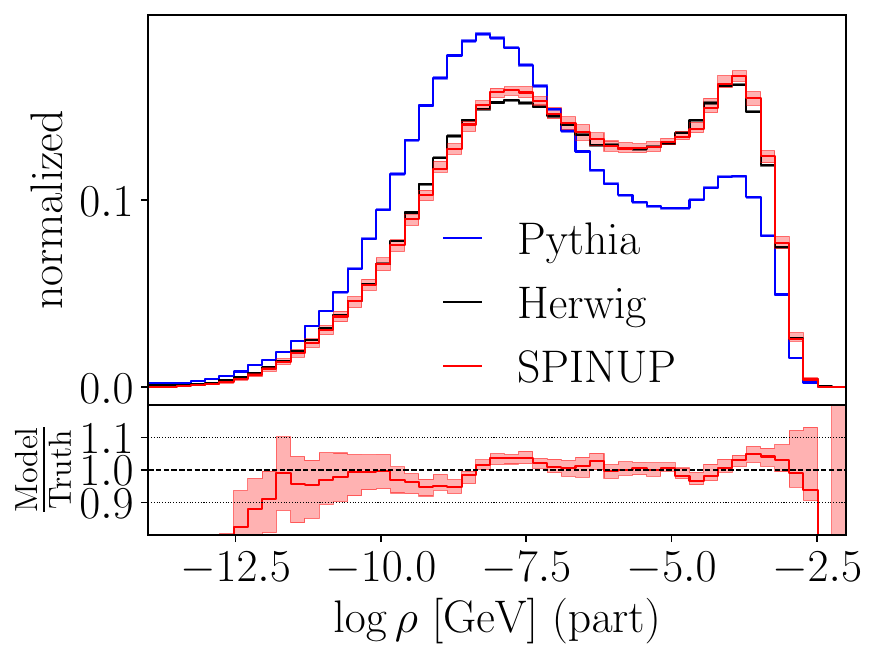}
    \includegraphics[width=0.32\textwidth, page=5]{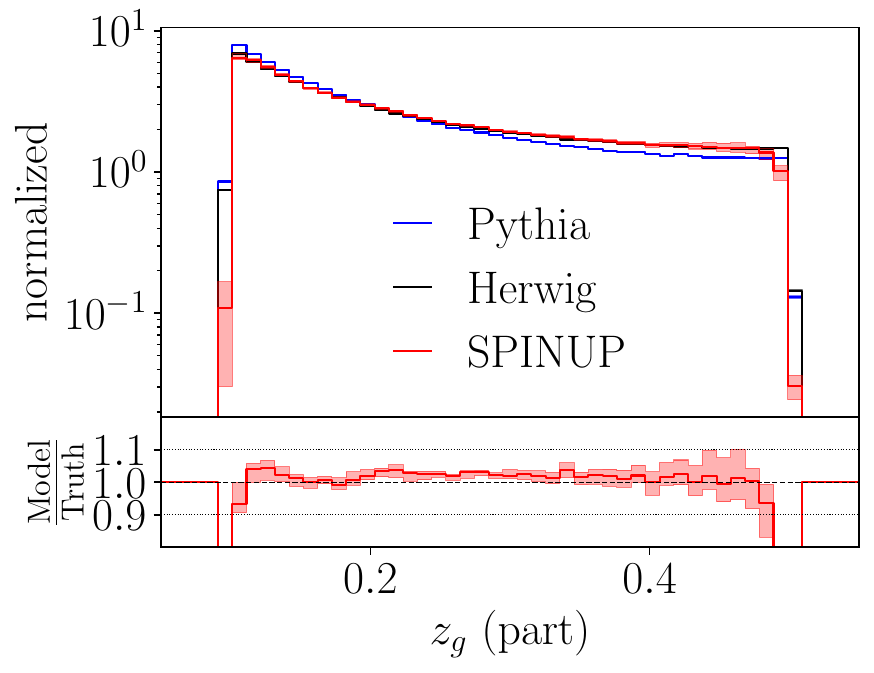}
    \includegraphics[width=0.32\textwidth, page=6]{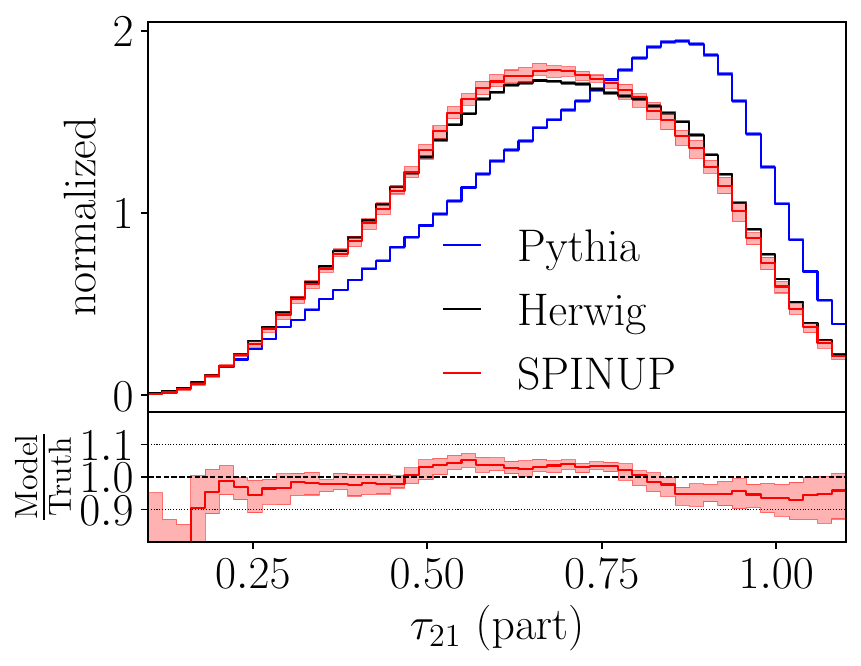}
    \caption{Part-level unfolded distributions obtained by running the method on the reco-level Herwig distribution.}
    \label{fig:omnifold_part_trueHerwig}
\end{figure}



\clearpage
\subsection*{Narrow features}

\begin{figure}
    \centering
    \includegraphics[width=0.49\textwidth,page=1]{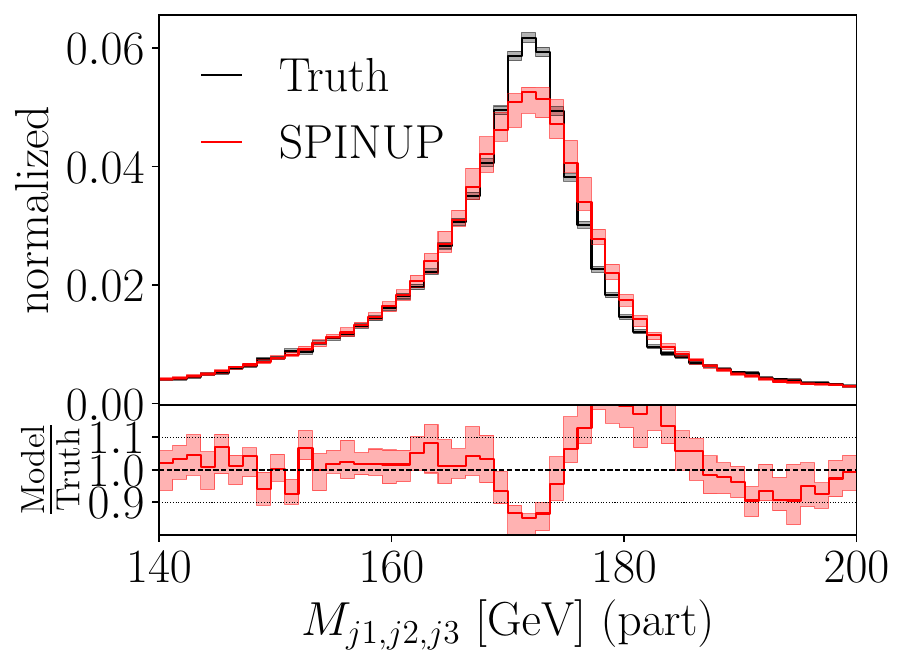}
    \includegraphics[width=0.49\textwidth,page=2]{figs/cms/observables_cms_final.pdf}
    \caption{SPINUP results for the $M_{j1,j2,j3}$ distribution at both part- and reco-level. Unfolding network trained to unfold the $m_t=169.5$GeV sample (Data).}
    \label{fig:cms_unfold_2d}
\end{figure}

A particular challenge for any unfolding algorithm is the reconstruction of narrow features that have been smeared out by the forward model. A common example of this is a sharp mass resonance. We illustrate this using the top decay dataset introduced in Ref.~\cite{Favaro:2025psi}. It consists of semi-leptonic $t\bar{t}$ decays 
\begin{equation}
pp \to t\bar{t} \to (bqq')\,(b\ell\nu) + c.c. \quad \text{with } \ell = e, \mu, \tau \; .
\end{equation}
The hard-scattering events are generated with Madgraph~\cite{Alwall:2014hca}, parton shower and hadronization are simulated with Pythia~\cite{Sjostrand:2014zea} and finally the detector response with Delphes~\cite{deFavereau:2013fsa}. Jets are clustered with XCone~\cite{Stewart_2015}, both on particle-level and on detector-level. A more detailed description of the simulation and the following event selection is given in Ref.~\cite{Favaro:2025psi}. \\
The goal is to unfold the kinematics of the three subjets originating from the hadronic top-decay. The reconstructed 3-subjet mass $M_{j1,j2,j3}$ then allows a measurement of the top mass. Past studies~\cite{CMS:2022kqg} identified the choice of top mass used in the simulation as the leading source of systematic uncertainty. 
\\
We train our transfer and NIS networks on a simulated sample with a top mass $m_t = 172.5$GeV. We then train SPINUP to unfold a sample with $m_t=169.5$GeV, the results are shown in Fig.~\ref{fig:cms_unfold_2d}. The left plot shows the results on part-level, here we observe that the model learns to recover the correct position for the triple-jet-mass mass peak, but the peak is significantly smeared out compared to the truth distribution. At reco-level, shown in the right plot, SPINUP perfectly recovers the measured data distribution. \\
This indicates that the method does not struggle with any prior dependence from the simulation top mass, but it does struggle with the degeneracy of the solution space. Due to the large smearing, a wide range of part-level distributions will give rise to almost indistinguishable reco-level distributions. The SPINUP ensemble does not map out the entire space of possible solutions here and underestimates the uncertainty. We attribute this to a preference of neural networks to learn smoother, less peaked functions. \\
This illustrate a general trade-off in unfolding problems. The SPINUP method is trained via likelihood maximization of the observed reco-data, without imposing any prior on the unfolded distribution. While this is generally a desirable property, in some unfolding problems, such as top mass peaks, we have a strong physics-inspired prior what the distribution should look like. We leave the investigation how to include such knowledge into the method for future work. 

\clearpage
\bibliography{tilman,refs}
\end{document}